\documentclass[12pt,preprint]{aastex}

\begin{document} 
\title{HST Photometry for the Halo Stars in the Leo Elliptical NGC 3377\footnote{
Based on observations made with the NASA/ESA Hubble Space Telescope, 
obtained at the Space Telescope Science Institute, which is 
operated by the Association of Universities for Research in Astronomy, Inc., 
under NASA contract NAS 5-26555. These observations are associated with program \#9811.
Support for this work was provided in part by NASA through grant 
number HST-GO-09811.01-A from the Space 
Telescope Science Institute, which is operated by the Association of 
Universities for Research in Astronomy, Inc., under NASA contract NAS 5-26555.}
}

\author{William E. Harris} 
\affil{Department of Physics \& Astronomy, McMaster University, Hamilton L8S 4M1, Canada}
\email{harris@physics.mcmaster.ca} 

\author{Gretchen L.~H.~Harris}
\affil{Department of Physics \& Astronomy, University of Waterloo, Waterloo N2L 3G1, Canada}
\email{glharris@astro.uwaterloo.ca}

\author{Andrew C. Layden}
\affil{Department of Physics and Astronomy, Bowling Green State University, 104 Overman Hall, Bowling Green, OH 43403}
\email{layden@baade.bgsu.edu} 

\author{Peter B. Stetson}
\affil{Dominion Astrophysical Observatory, Herzberg Institute of Astrophysics, National Research Council, 
5071 West Saanich Road, Victoria, V9E 2E7, Canada}
\email{peter.stetson@nrc-cnrc.gc.ca}

\shorttitle{Halo Stars in NGC 3377}
\shortauthors{Harris et al.}

\begin{abstract} 
We have used the ACS camera on HST to obtain $(V,I)$ photometry for 57,000
red-giant stars in the halo of the Leo elliptical NGC 3377, an intermediate-luminosity 
elliptical.  We use this sample of stars
to derive the metallicity distribution function (MDF) for its halo
field stars, and comment
on its chemical evolution history compared with both larger and smaller 
E galaxies.  Our ACS/WFC field spans a radial range extending
from 4 to 18 kpc projected distance from the center of NGC 3377 and thus
covers a significant portion of this galaxy's halo.  We find that the
MDF is broad, reaching a peak at log $(Z/Z_{\odot}) \simeq -0.6$, but containing
virtually no stars more metal-poor than log $(Z/Z_{\odot}) = -1.5$.
It may, in addition, have relatively few stars
more metal-rich than log $(Z/Z_{\odot}) = -0.3$, although 
interpretation of the high-metallicity
end of the MDF is limited by photometric completeness that affects
the detection of the reddest, most metal-rich stars.  NGC 3377
appears to have an enrichment history intermediate
between those of normal dwarf ellipticals and the much larger giants.
As yet, we find no clear evidence that the halo of NGC 3377 contains a 
significant population of ``young'' ($< 3$ Gy) stars.
\end{abstract}


\section{Introduction} 
\label{intro}

The imaging cameras on the HST have provided powerful tools to study large photometric
samples of individual stars in nearby galaxies that were quite out of reach
from ground-based instruments.  Although even the best photometry cannot provide the
level of detail that is achievable by high-resolution spectroscopy, 
these deep stellar samples open a valuable
route to understanding the early chemical evolution of their host
galaxies, compared with the much cruder information from integrated-light
photometry and spectroscopy.  

Among the possible target galaxies within
reach of the HST, the ellipticals are of special interest because
they may result from the widest possible range of formation histories,
from hierarchical merging at very early times, to recent major mergers,
to later growth by satellite accretion.  Representative dwarf ellipticals
are readily accessible within the Local Group (particularly NGC 147, 185, 205),
but the Local Group gives us no examples of giant or intermediate-sized E galaxies
to work with.
In a series of previous papers \citep{h99,h00,h02,rej05}, we presented analyses
of samples of stars in the closest giant elliptical, NGC 5128. These studies cover
four locations in its halo at projected distances ranging from 8 to 40 kpc.  In all
four of these studies the red-giant stars are clearly resolved, and in the
deepest one \citep{rej05}, even the old horizontal-branch population is reached.  We found
that in all four locations in the halo of this giant elliptical, 
the metallicity distribution of the giant stars
is clearly metal-rich (with $\langle$ Fe/H $\rangle \simeq -0.4$) and broad,
but with extremely small proportions of classically metal-poor stars (i.e. 
those with [Fe/H] $< -1$).  

Generalizing the results for NGC 5128 to all large ellipticals
may, however, be compromised by the possibility that NGC 5128 could
be a major merger remnant in which the majority of its halo stars were actually
formed within the metal-rich disks of its progenitor galaxies.  In fact, both
a major-merger or hierarchical-merging approach are capable of creating
an MDF in the halo with the same basic characteristics as we observe
\citep{bekki03,bea03}.  To extend the range of information
we have to work with, clearly it is of great interest to obtain
the metallicity distribution function (MDF), {\sl based directly on samples
of individual stars}, in other galaxies of the widest possible range of
properties.

NGC 5128, at a distance of only 4 Mpc, is by far the closest easily
accessible big elliptical.  To reach other ellipticals, 
we must step to the Leo group objects at $D \sim 10$ Mpc, and then
outward to Virgo ($D \sim 16$ Mpc) and Fornax ($D \sim 19$ Mpc), which
have the nearest accessible large samples of target galaxies.
The technological gains afforded by the step from the HST
WFPC2 camera to the newer ACS, with its much higher sensitivity, spatial
resolution, and field of view, now bring these more distant targets within reach
(see, for example, the recent photometric work of \cite{will06} to
resolve the brightest $\sim 1.5$ mag of the red-giant branches in four
globular clusters at the Virgo distance).
Roughly speaking, observing the halo of NGC 5128 with WFPC2 is comparable
(for the same number of HST orbits) with observing the Leo group ellipticals
with ACS.  

In this paper, we present new color-magnitude photometry for the halo stars in
the intermediate-sized E5 elliptical NGC 3377 in
the Leo group, and briefly compare its metallicity distribution function (MDF) with
those of both a dwarf and a giant.
NGC 3377 has a luminosity of $M_V^T \simeq -19.9$ falling 
between those of the Local Group dwarfs and the giant NGC 5128 that
have been observed in previous studies; it therefore provides an
attractive bridge  in the parameter space
of galaxy properties that we can add to the discussion of chemical evolution
models.
The key parameters for the galaxy are summarized in Table 1.

\section{Observations and Data Reduction}
\label{observations}

The imaging data we use for this study were obtained in our HST program 9811.  NGC 3377
is at relatively high galactic latitude and
low foreground reddening ($b = 58\fdg3$, $E_{B-V}$ = 0.03) and is thus
well placed for deep photometry of its halo stars, as free from
field contamination and foreground absorption as we can hope to have.

We used the Advanced Camera for Surveys in its Wide Field Channel,
with image scale $0\farcs05$ per pixel, to reach
the maximum photometric depth and field coverage.  A single field was
targeted at $\alpha = 10^h 47^m 49\fm00, \delta = +13\fdg 55\farcm 40\farcs0$
(J2000).  This field is $3\farcm82$ southeast of the galaxy center, equivalent to
$\simeq$12 kpc projected linear radius.    Since the effective radius of
the galaxy light profile is $R_e = 1\farcm1$, our field is well beyond the
central bulge and can plausibly be considered as giving us a first look at the genuine halo.
Fields further in would obviously be of interest to track any metallicity
gradient that the system might have, but would also be much more difficult
to measure because of considerably increased crowding.

The filters used were the ACS/WFC ``wide V'' ($F606W$) and ``wide I'' ($F814W$),
the same ones as in our previous NGC 5128 studies.   
The $(V-I)$ color index is a very useful metallicity indicator particularly for
old red-giant stars in the range [Fe/H] $\sim -1.5$ to $\sim 0$, which
as will be seen below, is our main range
of interest.  Total exposure times were 38500 sec over 15 full-orbit exposures in $F606W$,
and 22260 sec over 9 full-orbit exposures in $F814W$, adding up to a total of 24 orbits
on this single field.  The individual exposures were dithered over steps from 0
to 20 pixels, allowing elimination of most cosmic rays, bad pixels, and other artifacts
on the detector.  To prepare the images for photometry, we extracted the drizzled
individual images from the HST Data Archive, registered them to within 0.05 pixels,
and median-combined them.  This procedure gave us a single very deep 
exposure in each filter.  In Figure 1, we show the combined $I-$band image, demonstrating
that the halo
is well resolved into stars throughout.  A small inset portion of the field
is shown in Figure 2.  Many faint background galaxies can be seen through the
halo, but the overwhelming majority of detected objects in this field are
the red giant stars of NGC 3377 itself.  

To carry out the photometry on the two combined frames we used the standalone version of {\sl DAOPHOT} in
its latest ({\sl daophot 4}) version as written by one of us (PBS).  The 
normal sequence of {\sl find/phot/allstar}
was used, with point spread functions defined from the average of 50 to 60 stars spread
across the field.  The pixel coordinate systems on the two images were registered, and then
the detected stars on each one were matched up to find those measured in both colors.
The number density of stars on the frame was reasonably high (the average separation
between detected stars is $\simeq 7$ px, while the FWHM of the point spread function is
2.3 px or $0\farcs115$); also,
differences in magnitude are not useful for deciding on matchup of stars between
the $V$ and $I$ frames  since the target stars have
a very large true range in color. Matchup was therefore
done by coordinates alone in a careful iterative procedure:  five iterations were done with successively
increasing matching radius, starting with $\Delta r = 0.2$ px and increasing to 1.5 px.
In practice, more than 95\% of the 75000 matches between
the $V$ and $I$ images fell well
within 1 pixel of each other. 

The next step was to define the cleanest possible subset of these matches.
We rejected any detected objects that fell in the regions of 5 bigger-than-average
background galaxies and one bright star, simply by masking out those regions. 
We also rejected objects with very poor goodness-of-fit
to the PSF ($\chi_V > 2.5, \chi_I > 2.0$); and any with much larger than normal {\sl ALLSTAR}
magnitude uncertainties ($e_V, e_I > 0.25$). 

Although there is a very noticeable
gradient in the density of stars from the lower right portion of the field to
the upper left (see Fig.1), none of the areas are excessively crowded to the
point where photometry via normal PSF fitting is unusually difficult.
Our data do not enter the high-crowding regime that can be found in other
examples in the literature, such as (among others) the inner halo of
NGC 5128 \citep{h02}, the bulge of the Local Group dwarf M32 \citep{gri96},
or the halo of the Virgo dE VCC1104 \citep{har98}.
Nevertheless, to extract a conservatively ``uncrowded'' sample of stars
to work with, we calculated the nearest-neighbor distance (NND) for every measured
star on both frames.  We define the NND for a star simply as the distance
to the closest neighboring star on the list \emph{regardless of relative
brightness}.  The distribution of the NND values is shown
in Figure \ref{crowding}, showing that the typical separation between objects
is $\sim 5$ px or about twice the FWHM. 
To cull the photometry files of stars that are even mildly crowded, we decided to
reject \emph{any} object for which $NND < 3$ px, on either the $V$ or $I$ frames.
This step removed about 17\% of the $I$ detections and 12\% of the $V$ detections.
We regard the NND cutoff of 3 px as an extremely stringent choice; in fact,
\emph{allstar} has no difficulty handling cases down to $NND \sim FWHM$ or even
less. 

The final result of the photometry, after removal of (a) the masked regions,
(b) objects with high $\chi$ and high magnitude uncertainties, and (c) stars
for which any concern existed about crowding, left a total of 57039 well measured,
uncrowded stars. 

For calibration of the photometry, we chose to transform the filter-based
magnitudes $F606W, F814W$ into $V, I$ in order to facilitate comparisons with previous work
\citep{h99,h00,h02,rej05}.  First, we used the {\sl daophot/substar}
routine to subtract all but the brighter stars from the final averaged pair of
images.  We then obtained aperture photometry of these now-isolated bright stars
to correct the {\sl allstar} instrumental magnitudes to aperture magnitudes,
and extrapolated these to large radius following the standardized prescriptions
of \citet{sir05}. These gave us the magnitudes $F606W$ and $F814W$ on the 
natural ACS VEGAMAG filter system.  Finally, we independently derived transformations
of these to the standard $VI$ system by measuring an extensive set of images
of the NGC 2419 standard field from the HST Archive, and comparing these
with ground-based standard data in the same field.
The ground-based data are part of PBS's ongoing programme 
\citep{ste00} to maintain and upgrade an all-sky system of photometric
standards on the {\it BVRI\/} system of \citet{lan92}.  The
photometry of the NGC 2419 field in particular is discussed by 
\citet{ste05}.  This remote globular cluster has the advantages
(for calibration purposes) of a blue
horizontal branch at V$\,\sim\,$21, a red giant branch tip at ($V$, {\it
V-I\/}) = (17.3, 1.50), and a wealth of foreground stars.  The 1,257
stars adopted as standards in this field have a median of 46 groundbased
observations per star in each of the $V$ and $I$ filters, a
median standard error of 0.0054$\,$mag in $V$ and 0.0058$\,$mag in $I$,
and a magnitude and color range $ 12.66 < V < 22.64$, $-0.05 < (V-I) < 3.04$.
Keeping quadratic terms in color index, we found for these

\begin{eqnarray}
F435W \, = \, B + 0.135 (B-V) - 0.44 (B-V)^2 \\
F606W \, = \, V - 0.265 (V-I) + 0.025 (V-I)^2 \\
F814W \, = \, I + 0.028 (V-I) - 0.008 (V-I)^2
\end{eqnarray}

\noindent These equations reproduce the NGC 2419 standard stars in all bands
to within a scatter $\sigma = \pm 0.03$ mag.  
The equation for $B$ is not used in this paper but is listed for information.
These transformations are very close to those published in \citet{sir05},
and also to those derived for our NGC 5128 outer-halo field \citep{rej05}, which
were taken during the same Cycle.
The final color-magnitude diagram for the complete sample of
57039 stars is shown in Figure \ref{cmd1}
and will be discussed below. 

We estimated the internal photometric uncertainties and the detection
completeness through a series of artificial-star tests with the {\sl addstar} component
of {\sl DAOPHOT}.  Stars were added to the combined $V$ and $I$ images in groups of
1000 over a wide range of magnitudes; these experiments
were done independently on the $F606W$ and $F814W$ images.  
The images were remeasured in the same way
as the original frames.  The fraction $f$ of stars recovered, as a function of instrumental magnitude,
is shown in Figure \ref{completeness}.  
(The magnitudes here are the filter-based ones $F606W$ and $F814W$,
discussed below.)  The limits of our data, defined as the magnitudes at
which $f$ drops to 0.5, are $F606W(lim) = 28.95$ and $F814W(lim) = 27.70$.  The trend of
$f$ with magnitude is well described in each case by a Pritchet interpolation function
\citep{fl95}, an analytic function with two parameters: the limiting magnitude,
and a parameter $\alpha$ giving the steepness of the dropoff.  For
these images we find $\alpha=2.5$ for $V$ and $\alpha=2.7$ for $I$.
Because of the color terms in the transformations (Eqs. 2,3), these limits
in the native filter-defined magnitudes do not correspond to single 
$V$ or $I$ values, but for a giant star with a typical color of $(V-I) \simeq 2$,
the limits are $V(lim) \simeq 29.4$ and $I(lim) \simeq 27.7$.

In the color-magnitude diagram of Fig.~\ref{cmd1}, the 50\% completeness lines are shown.
An important feature of these lines is that the limiting
curve for $V$ (the upward-slanting line on the right side of the CMD) cuts off
our ability to see any extremely red stars that might actually be present;
these would fall at the most metal-rich end of our metallicity distribution
function.  Considerably deeper exposures in $V$ will be needed to explore the true
``red limit'' of the giant stars in this galaxy.  Within the limits imposed by
the photometry, we explicitly take into account the completeness fraction $f$ in
our derivation (below) of the metallicity distribution.  As will be seen later,
the completeness cutoff may affect how much of the metal-rich end of the MDF is
ultimately detectable.

We find that the 50\% completeness level does not change significantly
with radius $R$ from galaxy center, except perhaps marginally in the region $R < 2'$
of highest crowding.  
A visual confirmation is shown in Figure \ref{4panel_cmd},
where we subdivide the CMD data into four radial regions.  The placements
of the completeness cutoffs do not change relative to the distribution
of stars (we will further justify this point in the later discussion
on the radial gradient of the metallicity distribution).
A further indication of the relative importance of crowding is shown in
Figure \ref{skylevel}, where we show the local sky intensity around
each measured star (in digital units,
directly as returned by {\sl allstar}) as a function of radius from the
center of NGC 3377.  The inward increase becomes
steeper within $R = 2'$, but the net change over the entire frame is
quite modest.
However, in the discussion that follows (for the distance measurement and the
metallicity distribution function), to remove any residual concerns
about crowding issues we further restrict our analysis
to the ``safest'' region $R > 2'$.

The random uncertainties of the photometry, as derived from the artificial-star tests,
are shown in Figure \ref{random_errors}.  
The rms uncertainty rises smoothly with magnitude in a roughly
exponential manner, shown by the curves in the Figure.  These have equations
$\sigma(V) = 0.05$ exp$((F606W-26.00)/1.8)$ and $\sigma(I) = 0.05$ exp$((F814W-25.5)/1.3)$.
In the analysis, we make no use of the data fainter than the 50\% completeness level
(marked by the vertical line in the figures).
Finally, in Figure \ref{photom_errors} we show the mean trends for systematic bias in the photometry:
a sample of the artificial-star tests is plotted, showing the median difference
$\Delta m = (input - measured)$ for both filters.  
The median lines, plotted in 0.1-mag bins, show a consistent trend 
for stars to be measured $0.01 - 0.02$ mag
too bright, a common feature of photometry of faint stars
within moderately crowded fields.  Fainter than the 50\% completeness level, the
mean curves diverge strongly to positive $\Delta m$ and the photometry becomes
systematically unreliable.  It is important to note,
however, that for the stars brighter than the completeness limit, the
mean biasses in the two filters run closely parallel, and the net bias in mean
{\sl color} $(V-I)$ -- an important consideration since the color index is our
main metallicity indicator -- is negligibly small
and has no effect on the derivation of the MDF.  In addition, the random
uncertainties have no significant effect on the inferred properties of the MDF,
because we use only the brightest $\simeq 1$ mag of the RGB where the intrinsic
color range of the stars is far larger than the photometric scatter.  At the
very top of the RGB, the intrinsic spread is $\Delta(V-I) \simeq 1.6$ mag
while the photometric scatter is $\pm$0.1 mag or less.

\section{Distance Calibration}

NGC 3377 is part of the Leo I group along with several other large
galaxies, most of which are spiral-type.  Numerous measurements of distance
to these individual Leo members are in the literature from a variety of
well established distance indicators including Cepheids, planetary nebula
luminosity function (PNLF), surface brightness fluctuation (SBF), and
the tip of the old-red-giant branch (TRGB).  

\noindent {\sl NGC 3377:}  For NGC 3377 itself, the SBF
method \citep{tonry01} gives $(m-M)_0 \equiv \mu = 30.25$, while the PNLF method
\citep{ciar89} gives $\mu = 30.07 \pm 0.18$.

\noindent {\sl NGC 3379:}  For the single giant elliptical in Leo I, the TRGB method
has been applied through HST photometry in both the optical $I$ band and the
near-infrared.  The former \citep{sakai97} gives $\mu = 30.30 \pm 0.27$, and the
latter \citep{gregg04} gives $\mu = 30.17 \pm 0.12$.  The PNLF method \citep{ciar89} gives
$\mu = 29.96 \pm 0.16$, and the SBF method \citep{tonry01} $\mu=30.12$.

\noindent {\sl Large Spirals:}  Cepheid-based distances have been published
for some of the major spirals in the group.  These include $\mu = 30.01 \pm 0.19$ for 
NGC 3351 \citep{graham97}, $30.25 \pm 0.18$ for NGC 3368 \citep{tanvir99}, and
$30.10 \pm 0.14$ \citep[unweighted average of 27 Cepheids;][]{saha99} or
$29.71 \pm 0.08$ \citep{freedman01} for NGC 3627.  PNLF distances for these
galaxies include $\mu=30.05 \pm 0.16$ for NGC 3351 and $29.99 \pm 0.08$
for NGC 3627 \citep{ciar02}.

\noindent Treating all these Leo members as if they are at the same true distance
from us and taking an indiscriminate average of all these measurements suggests
a rough consensus near $\mu \simeq 30.1 \pm 0.05$, or $D \simeq 10.4$ Mpc
for the group as a whole.  The galaxy-to-galaxy dispersion of these measurements,
which is $\sigma_{\mu} = 0.17$ mag, is quite similar to the typical internal uncertainties
of each one and gives no strong evidence that the distance depth of the group
is an important factor.

Our new photometry penetrates well into the the old-halo red giant branch of NGC 3377,
with a large sample of stars,
and thus provides a new opportunity to use the TRGB distance indicator rather
precisely.  The key parameter is the ``tip magnitude'' which is,
physically, the luminosity of the helium flash in the core of the red giant
as it reaches the top of its first ascent along the giant branch.
The bolometric luminosity of the RGB tip is, fortunately, only mildly
dependent on metallicity for old stars, allowing it to be turned into an
accurate standard candle.

Our analysis follows the methods used in 
\citet{sakai96,sakai97} and \citet{h99}, among others:  we plot up
the luminosity function of the RGB stars and use the detailed shape of the LF
in the $I$ band to define the onset of the RGB.  For stars more metal-poor than
[Fe/H] $\simeq -0.7$ (which include the majority of the ones we measure here;
see next section), the $I$ band has the strong advantage that the differential
bolometric correction across the top of the RGB is almost cancelled by
the dependence of $M_{bol}(tip)$ on metallicity, leaving $M_I(tip)$ 
with only a gradual slope with increasing color.

The luminosity function is shown in Figure \ref{lf}, based on 49380 stars
beyond $R>2'$ from the galaxy center.  It has been 
smoothed with a Gaussian kernel of $\sigma_I = 0.02$ mag.  Completeness
corrections are quite unimportant here, since the $f=0.5$ completeness level
is $I \simeq 27.6$, much fainter than the well resolved top of the RGB.
The fact that the LF rises gradually upward at the TRGB (rather than
the ideal case of an abrupt jump) is a normal consequence of an intrinsically
steep LF at that point, convolved with the observational photometric scatter,
and adding in the smoothing effect of
the slight downward slope of the tip toward redder colors, and a small
amount of field contamination, all of which act to
blur out the TRGB point to some extent \citep[see][for detailed discussion and
methodology]{har98,h99}.
The actual tip is clearly somewhere near $I = 26$, but to define the true
TRGB precisely, we look for a strong 
change in the slope of the LF as we go to fainter
magnitudes across the tip.  The numerically calculated first derivative of
the LF, $dn/dI$, is shown in the second panel.  A sharp upturn is present
past $I = 26$ with well defined peaks that appear at $I = 26.2$ and 26.3.
These are independent of the precise value of the smoothing kernel within
broad limits.  To decide which of these to pick we use the ``edge response
filter'' or ERF (the numerical second derivative of the LF), plotted in
the lower panel of Fig.~\ref{lf}.  Here, we adopt the first and most prominent
peak at $I(tip)= 26.2 \pm 0.1$.  

The distance modulus follows immediately once we apply a fiducial value
for $M_I(tip)$.  The most well established recent calibration from within
the Milky Way is from a very large sample of stars in $\omega$ Cen
\citep{bell04}, giving $M_I(tip) = -4.05 \pm 0.12$.  Calibrations based
on theoretical RGB models \citep[e.g.][]{sal02} yield values
in the range $-3.95$ to $-4.22$ depending on both the details of the stellar
physics and on observational constraints, but are entirely consistent with
the $\omega$ Cen value, which we adopt here.  We therefore
obtain  $(m-M)_I = 30.25 \pm 0.15$ for NGC 3377.  This must be corrected for a foreground
absorption 
of $A_I = 0.07\pm0.02$, giving a final TRGB distance measurement $\mu = 30.18 \pm 0.16$.

Averaging the TRGB distance in with the SBF and PNLF measurements listed above,
and giving the three methods equal weights, we arrive
at an average $(m-M)_0 = 30.17 \pm 0.10$,
or $D = 10.8 \pm 0.5$ Mpc for NGC 3377 itself.  
These three methods -- all based, necessarily, on the
properties of the old stellar population -- are in excellent mutual agreement, and
we consider the distance to this system to be as well established as any in
the local galactic neighborhood.

\section{The Nature of the Brighter Stars}

In Figs.~\ref{cmd1} and \ref{4panel_cmd} there are numerous stars
scattered above the TRGB.  There are $N_{bright} \simeq 1360$ such
stars within $24 < I < 26$.  Could these be a signature of a younger,
intermediate-age population ($\tau \lesssim 3$ Gy), which would indicate some more
recent star formation in the galaxy's history?
in Figure \ref{xy2}, we show the spatial distribution of these stars
across our ACS field, compared with the distribution of the stars
in the brightest half-magnitude of the RGB itself.  Both types of
stars show an obvious gradient decreasing from galaxy center, and so
a significant fraction of $N_{bright}$ must be genuine members 
of NGC 3377.

Before placing limits on the number of such stars that might
genuinely be present, we need to rule out other possible contributors,
including (a) field contamination, (b) accidental blends of two normal RGB
stars, or (c) giants or AGB stars in highly evolved or temporary
states including long-period variables (LPVs), which can contribute
noticeable numbers of supra-TRGB stars even in an old population
\citep[see][for more extensive discussion and modelling]{rej03,gregg04}.  We dealt with a
similar issue for the halo stars in NGC 5128 \citep{h99}
and found that only $\sim$1\% of the population there could be ascribed
to the intermediate-age category.

The first option (field-contaminating objects) 
will include both foreground stars and faint,
very small background galaxies that are near-starlike in appearance.
The number of foreground stars in the direction of NGC 3377
and over the area of ACS/WFC should be $N \lesssim 20$ from
Galactic models \citep[e.g.][]{bah81}, but the number of faint,
misidentified background galaxies is almost certainly larger and
would best be measured from an adjacent `control' field, which we
do not have.  However, we can specify a reasonable upper limit to
$N_{field}$ by using the outermost regions of our measured field.
For $R_{GC} > 4\farcm5$ (an area of $2.78$ arcmin$^2$)
there are 94 objects within $24 < I < 26$, suggesting that over the
entire 11.33 arcmin$^2$ field $N_{field}$ should be at most 376.
The number density of stars is still declining at the outer
edges of the field, and so this number must be a generous upper limit.
Since we see 1360 such objects over the whole field, this suggests
$\sim 1000 - 1200$ of them are intrinsic to NGC 3377.

Some objects could appear in this bright range because of accidental
blends of two stars that are both near the upper end of the RGB.
Blends like this are
very likely to be responsible for much of the scatter of stars
appearing just above the nominal RGB tip\footnote{There are $\simeq 1300$
stars in our CMD in the small interval $I = 24.0 - 24.2$ just above the RGB tip
that are responsible for blurring out the definition of the TRGB
and creating the smooth rolloff in the luminosity function
just above the tip; see Figure 10.  Most of these objects
can be understood as due to blends of the brighter RGB stars with 
the huge number of RGB stars
$\gtrsim 1.5$ mag {\sl fainter} than the tip.},
but not for the ones $\sim 0.7$ mag or more above it.
Statistically the number of blended pairs will increase as the
square of the number of stars per unit area,
\begin{equation}
N_{blend} \, \simeq \, {N_{\star}^2 \over 2} {{\pi q^2} \over d^2}
\end{equation}
where $N_{\star}$ is the number of stars on the frame capable of
generating a blended pair brighter than the TRGB, $q$ is the radius
of one resolution element, and $d^2$ is the area of the field.
Adopting $q \simeq 2$ px, and using the fact that
there are $\simeq 30000$ stars within 1 magnitude of the RGB tip,
we expect $N_{blend} \sim 350$, or about one-quarter of all
the objects appearing above the TRGB.  

The presence of LPV-type stars is favored in an intrinsically
metal-rich old population \citep[see][]{rej03}, 
and their numbers should be proportional to the total
luminosity of the whole population contained in our field.
Extrapolating from the stars brighter than our completeness limit
with a standard Population II luminosity function, we estimate
very roughly that $V(int) = 14.7$ or $L \sim 8 \times 10^7 L_{\odot}$.
From \cite{ren98} we then find $N_{LPV} \simeq 400$, similar to
the expected number of accidental blends.  An observational confirmation
that LPV-type variables should indeed be there to be found is discussed
by \citet{gregg04}, who present $HST/NICMOS$ near-infrared photometry
of the RGB stars in inner-halo fields for NGC 3379, the other Leo
elliptical.  They find (see their Table 3) that between one-quarter
and one-half of the detected stars above the TRGB are 
variable, consistent with our rougher estimate.  \citet{rej03} found
a total of more than 1100 LPVs in two fields around NGC 5128 covering
an area of 10.46 arcmin$^2$, and state that these make up roughly
half of the total number of stars above the TRGB, again consistent with the estimates
made here.

In summary, we suggest that more than half the population of stars
that lie clearly above the RGB tip are due to 
a combination of blends, temporary high-luminosity
states of the giants, and a small amount of field contamination.
The remainder not accounted for by these effects is thus
$N_{bright} \sim 400 \pm 100$, or $\simeq$1\% of our observed
total on the field.  We conclude tentatively that there is no evidence
for any significant ``young'' ($\tau < 3$ Gy) population of stars
in our field.

These rough estimates are not intended to replace a more comprehensive
population synthesis analysis, based on simulating the observed RGB
with stars drawn from model isochrones over a complete range of
ages and metallicities \citep[see, e.g.][for a recent example]{will06}.
In a later stage of this work we will develop such simulations and
constrain the age range of the halo stars more completely.

\section{The Metallicity Distribution}

\subsection{Deriving the MDF}

We are now in a position to use the colors of the RGB stars to gain a
first direct measurement of the metallicity distribution function (MDF) in this
galaxy.  To enable direct comparisons with other systems, we follow the
same method as used previously for NGC 5128 and selected dwarf ellipticals
\citep{h99,h00,h02}.  A finely spaced grid of RGB evolutionary tracks
is superimposed on the measured CMD, and we interpolate between these to
tag every star with a heavy-element abundance $Z$ based on its location
within the tracks.  As before, we use the $\alpha-$enhanced tracks of
\citet{van00}, and calibrate them in terms of absolute $(V-I)$ color
by force-fitting them to the observed RGB sequences for Milky Way globular
clusters.  This approach is described in detail in \citet{h02} and we 
do not repeat it here. However, we
emphasize that the metallicity scale we derive this way is 
an {\sl observationally calibrated one} based on real globular clusters.
{\sl The theoretical RGB tracks are used only to aid interpolation between
the observed sequences for real clusters,} and thus our results
do not depend critically on the particular choice of stellar models.
The one unavoidable and model-dependent assumption that underlies this
method is that the mean age of the NGC 3377 stars is taken to be the
same as for our local globular clusters (that is, about 12 Gy).  If in
fact they are systematically younger, then this method would slightly
underestimate their $Z-$abundances since the RGB locus shifts blueward
at lower age.  However, as long as the age differences are not severe,
the age correction to the MDF is small; for example, in \citet{rej05} we note that
a shift in mean age from 12 Gy to 8 Gy would produce only a $0.1-$dex underestimate
in log $Z$.  See also the RGB track comparisons for different ages
in \citet{h99}, which demonstrate similar conclusions.

In Figure \ref{cmd_fiducial} we show again the composite color-magnitude diagram,
but now with the RGB tracks added.  The solid lines are those of the
\citet{van00} grid, whereas the two dashed lines at right are ones
for Solar ($Z=Z_{\odot}$) and $\simeq 3 Z_{\odot}$ metallicities.  
It is clear, as we noted above, that the photometric detection limit for
red stars set by the $F606W$ exposures would eliminate any giant stars
at Solar metallicity or above from our sample, even if they were present.
What is not obvious from this Figure is whether or not this detection cutoff
imposes any more basic limit on our interpretation of the MDF; that is,
are many high$-Z$ stars actually likely to be present in this galaxy?
To gain a better reply to this question, we need to look at the details
of the MDF itself as far as we can gauge them.

Our derived MDF is shown in Figure \ref{feh_3panel},
where we divide the sample into half-magnitude bins by approximate
luminosity $M_{bol}$.  By doing this, we test for any systematic
errors in the interpolation procedure that might result from incorrect
placement of the RGB model grid on the data; we also use it to 
reveal any broadening of the deduced MDF from photometric scatter
toward the faint end of the data.  Encouragingly, we see no systematic
shift in the shape of the MDF:  in all three bins, the peak occurs
at log $(Z/Z_{\odot}) \simeq -0.6$, with a broad tail extending to
lower metallicity and a much steeper ramp-down to higher metallicity.
As we found for the giant elliptical NGC 5128 \citep{h02}, remarkably
few stars are more metal-poor than log $(Z/Z_{\odot}) = -1.5$.
As a visual confirmation of this point, note directly from the CMD
how few stars there are near the bluest RGB tracks, particularly at
the high-luminosity end where the track color is most sensitive to
metallicity.  In other words,
NGC 3377 -- {\sl like the giant ellipticals, but unlike the smallest dwarfs
and globular clusters} --
contains few stars indeed that are in the same
regime as the ``classic metal-poor'' halo of the Milky Way.
This comparison (see also the discussion below) indicates that
NGC 3377 did not form simply by the amalgamation of fully-formed
dwarfs.

Fig.~\ref{feh_3panel} explicitly shows the MDF with, and without,
photometric completeness corrections, since these are important
to our assessment of the MDF.  Any extremely red 
stars falling to the right of the $f=$50\% line in the CMD have been
rejected from the sample, since in this region the completeness
correction itself becomes dangerously large and the random and
systematic errors of the photometry increase rapidly.  
For all stars brighter than this cutoff level, the completeness-corrected
samples (the open histograms in Fig.~\ref{feh_3panel}) have been constructed
by weighting each star individually as $(1/f)$ where $f = f_I \cdot f_V$
is the combined completeness fraction at its particular location in
the CMD.  For comparison, the unweighted MDF (based only on counting up
all stars with $f > 0.5$) is shown in the hatched regions.

Because the 50\% completeness line in $V$ runs nearly parallel to the
RGB tracks in its region, it essentially has the effect of cutting off the MDF
rather abruptly at log $(Z/Z_{\odot}) = -0.3$.  Redward of this point,
we have no reliable knowledge of the MDF shape.  
But by the time we reach this photometrically-driven cutoff, the MDF is
already well past its peak frequency and declining steeply.  The evidence
is thus strongly suggestive that we have seen the majority of the 
total MDF, and that only a small fraction of its stars are at Solar-type
abundance or higher.  By contrast, the giant NGC 5128 has an MDF extending
well up past Solar abundance, even in the outer reaches of its halo \citep{h02,rej05}.

The fact that the $V$ completeness cutoff is so close to the MDF peak, however,
leaves lingering doubts that we have indeed seen the full shape of
the metallicity distribution.  Could there, in fact, be significant
numbers of more metal-rich stars in the galaxy that perhaps belonged to
a later star-forming episode and are still waiting to be identified?
Such stars would be almost as bright in $I$ as the other RGB stars, but
would have been too red to have been detected in $V$.  To look into
this a bit further, we have used the
$I-$band measurements alone to isolate
stars that were (a) detected and measured on the $I-$band image; 
(b) {\sl not} detected on the $V-$band image; and (c) comparably bright
in $I$ to the other RGB stars.  
If there are large numbers of these, it would be suggestive that
we have missed much of the actual RGB.  

The luminosity function for the stars measured {\sl only} in $I$, and passing
the same photometric culling described above, is shown 
as the dotted line in Figure \ref{2lf}.  
This LF rises to a peak at the
completeness limit $I \simeq 27.7$ and then falls steeply.
By contrast, the LF for the stars measured in {\sl both} $V$ 
and $I$ (solid line in Fig.~\ref{2lf}, replotted from Fig.~\ref{lf}) 
has a broad peak determined by the $V-$band completeness limit.  

Very metal-rich RGB stars (those with log $Z/Z_{\odot} > -0.3$)
would belong to the ``$I$ only'' LF and would
lie in the approximate range from $I \simeq 26.6$ (for the very top of the RGB
at $(V-I) \sim 3$) down to the $I$ completeness limit.  
We see that the LF for the stars in that range (dotted line in the Figure) stays well
below the LF for the stars with both $V,I$ (solid line).
In the range $26.6 < I < 27.6$, 20008 objects were measured {\sl only}
in $I$ but not $V$. 
But this total should be considered as only an 
{\sl upper} limit to the actual number of metal-rich RGB stars, because
close visual inspection of the ``$I$ only'' candidates shows that a
high fraction of them are small clumps of bad pixels, or faint, very
small background galaxies.  These passed the $\chi$ and $e_I$
selection steps, but did not match objects appearing
at the same place on the $V$ image and thus were rejected in the 
construction of the CMD.  
Taking these into account, we make a {\sl very rough} estimate 
that there are fewer than $\sim 10000$ genuine RGB stars redder than the 
completeness cutoff in
our CMD.  By comparison, if we employ a simple chemical evolution
model to predict the shape of the entire MDF and use it to extrapolate
the observed MDF up past the photometric completeness cutoff (see next section), we
find that 10\% of the MDF should lie at $Z > 0.5 Z_{\odot}$,
corresponding to $\sim 5700$ `missed' stars in addition to our observed
sample of 57000 over all radii.
These very metal-rich stars would not change the MDF in a major way
if (for example) they formed only the high-metallicity tail of the MDF that
we already see.  Unfortunately,
any more definitive statement will have to await deeper photometry 
than is currently available.  A clear answer would require doubling
the $V$ exposure time; that is, adding another 15 orbits to what we already
have in hand.  The incremental cost of HST time would thus be significant.

\subsection{A Chemical Evolution Model}

To step beyond the raw MDF into a simple physical interpretation, we use
the basic chemical evolution model outlined in 
\citet{h02} (hereafter HH02) that we applied to
the NGC 5128 halo and bulge stars. Similar models have also
been used for the halo of the Milky Way
\citep{prantzos03}, and the globular cluster systems of large
galaxies \citep{van04}, among other situations.  This is the
so-called ``accreting box'' model in which we assume that a region
of primordial ($Z \simeq 0$) gas turns itself into stars through a long
succession of star-forming episodes, all the while that more gas is
flowing into the region.  The rate of gas infall is assumed to die away
gradually and smoothly with time, so that in the late stages of the
region's history, its chemical evolution asymptotically approaches
the classic ``closed box'' or ``simple'' model.  In a general sense, this
model is a first-order description of what would be expected to happen
during hierarchical merging of a large set of initial, pristine gas clouds,
wherein star formation within the clouds is happening simultaneously with
the merging into a bigger final galaxy.

In HH02 we postulate that the gas infall rate starts at a rather
high level and then dies away as an exponential decay with time. 
We also assume that at each star formation step, the same 
(small) fraction of ambient gas gets turned into stars.
The abundance $Z$ of the stars
forming at any given moment is then determined by the $Z(gas)$  
left behind by the previous steps, in addition to the amount and
composition of the 
new gas entering the region just before the next star formation
step occurs.  The sequence of timesteps can be numerically integrated
to give the final distribution $n(Z)$ of the stars (see HH02 for details).
The model in its simplest form has three essential 
and unavoidable parameters:  the effective
yield $y_{eff}$ of nucleosynthesis; the initial gas infall rate relative
to the amount of gas initially present in the region; and the
exponential decay time for the infall.  

An extremely effective way to exhibit the match between model and
data is through the linear form of the MDF, as number of stars per
unit heavy-element abundance $Z/Z_{\odot}$.  The key results are
shown in Figure \ref{zmodel}, where we include stars brighter than
$M_{bol} \simeq -2.5$, i.e. ones in the upper two panels of Fig.~\ref{feh_3panel}.  As before, we include only stars at radii $R_{GC} > 2'$.
The raw (shaded)
and completeness-corrected (unshaded) MDF are both shown. One
particular accreting-box model solution is shown as the curved line.  For
$Z \gtrsim 0.5 Z_{\odot}$, the observed MDF is cut off by photometric
incompleteness, but as noted above, this cutoff is well past 
the MDF peak at $Z \simeq 0.15 Z_{\odot}$, and the MDF is seen to
decline strongly up till the photometric cutoff.  

The model fit shown in Fig.~\ref{zmodel} is for a triad of parameters that
match the {\sl completeness-corrected} data.  The effective yield is
$y_{eff} = 0.0039 \simeq 0.26 Z_{\odot}$; the exponential decay time for infall
is 30 timesteps; and the total amount of gas infall is equal to 4 times the
initial gas present in the region.  The numerical integration is carried
through several hundred timesteps until almost no gas remains.
In each timestep we assume following HH02 that 5\% of the ambient gas is converted
to stars, and that the infalling gas has $Z=0$.
As we commented in HH02, the absolute
value of the timestep $\delta t$ is arbitrary in this model and must be supplied
by other methods; however, through various arguments about the actual rate of 
hierarchical merging (see again HH02), it is plausibly near $\delta t \sim $ 30 Myr give or
take factors of two.  If so, the main period where infall is strongly 
influencing the MDF shape is thus the first $1 -3$ Gyr.  

Each of the three critical parameters in the model
controls three distinctly different features of the final MDF.
First, the rapid rise in $n(Z)$ from $Z = 0$ to $Z = 0.1$ is set by the initial infall
rate:  the less the infall of pristine gas, the smaller the rise,
until in the limit of zero infall, the $n(Z)$ curve approaches the simple
exponential decline of the closed-box model, $n \sim e^{-Z/y}$.
Second, the location of the peak, where the MDF turns over and begins to decline,
is driven by the infall decay time: a longer decay time produces
a higher$-Z$ peak point since there is more gas to allow the enrichment
process to continue driving upward.  Finally, the long exponential tail to higher $Z$,
where infall has essentially stopped and the only process left is the conversion
of the remaining gas into stars (that is, a closed-box model), is shaped by
the effective yield $y_{eff}$.

We have restricted the present comparisons to the simplest set
of parameters possible within the context of the model, and the
three that are the least co-dependent.
But additional model parameters are possible:  for example, ones
introduced in our previous analysis of NGC 5128 included an initial, very short,
time period $\tau_1$ over which the gas infall is constant or rapidly ramps up;
and also the assumed abundance $Z_i$ of the infalling gas.  However, these prove to
be much less critical to the solution than the three primary parameters
defined above.  The MDF data do not have enough
resolution at extremely low $Z$ to constrain $\tau_1$;
and any plausibly small abundance for the infalling gas in the range
$Z_i \lesssim 0.1 Z_{\odot}$ can be used to generate adequate matches to the data
(see HH02 and Vandalfsen \& Harris 2004 for more extensive discussion).

The value of the effective nucleosynthetic yield at about one-quarter Solar
is a factor of two or three smaller than standard $y-$values that would hold in
a closed-box model where all the gas remains held {\sl in situ}
and where the stellar IMF and nucleosynthesis rate are normal
\citep[e.g.][]{pagel75}.  But if some fraction of the gas escapes in stellar
and SN-driven winds (leaky-box) then the effect is to lower $y_{eff}$
in proportion \citep{binney98}.   Formally, our results indicate that about
half the gas present in the NGC 3377 region was ejected during the
successive rounds of star formation, preventing the enrichment from
proceeding upward past about one-half Solar levels.  For comparison,
in HH02  we found $y_{eff} \simeq 0.3 Z_{\odot}$ for the outer halo of NGC 5128,
similar to the solution for NGC 3377.
For the inner region of NGC 5128, much deeper in its potential well
where virtually all the gas could be held, we found $y_{eff} \simeq 0.85 Z_{\odot}$.

For systems such as these, with broad MDFs and obviously extended star
formation histories, much more complex models can be clearly brought to bear.
Numerous examples of such models can be found in the
literature \citep[e.g.][among many recent discussions]{naab06,rom05,wor05,scan05,valle05,laf04,oey00}.
These have been developed all the way to full hierarchical-merging and 
chemodynamical-evolution codes \citep[e.g.][]{kawata03,cole00,bea03,som99}.
In some cases very different parametric approaches are used; for
example, \citet{wor05} advocate the use of a closed-box model in 
which $y_{eff}$ changes with time.
In such a model, quite a strong decrease in $y$ with increasing $Z$
is necessary to reproduce the rising part of the MDF at low $Z$.  In our view,
the use of gas infall combined with $y_{eff} \simeq const$ is much more
physically plausible, and has the strong advantage of employing a process
(inflow of low-metallicity gas) which is almost certainly happening anyway at early times.

It is notable, however, that the minimally simple set of parameters we use here
already provides an accurate first-order fit to the MDF.
The interpretation we offer for the validity of our
first-order model is simply that in galaxies this large, we
are looking at the combined results of so many star-forming events
that much of the individual detail has been averaged out and is now
no longer distinguishable.

\subsection{Gradients and Fine Structure?}

We can also use the MDF to search for any indication of a radial metallicity
gradient.  Our ACS/WFC field has stars with projected radial distances from the center
of NGC 3377 from $1\farcm3$ to $5\farcm7$ (4.1 kpc to 17.7 kpc), a large enough
range to be worth exploring for trends.  The MDF, again in its linear form, is shown
in Figure \ref{z_3panel} for an inner zone ($1\farcm3 - 2\farcm0$),
a middle zone ($2\farcm0 - 3\farcm5$), and an outer zone
($3\farcm5 - 5\farcm7$).  In each case, we show the {\sl same chemical
evolution model} applied to all three radial bins that we deduced for
the total population in Fig.~\ref{zmodel} above, simply renormalized
for the different total numbers of stars.  

All three panels are similar.  We see no evidence for
a radial gradient in metallicity, either in the peak position of the
MDF or in its detailed shape.  Although we deleted the innermost
region $R < 2'$ in the earlier analysis to minimize random scatter
in the CMD, LF, and MDF, it has not led to any systematic bias
in our understanding of the metallicity structure.  This is, perhaps, not a surprising
result since the effective radius of the galaxy is $1\farcm1$ and
thus all the stars we sample are well outside it.  In NGC 5128,
we found that the MDF is virtually unchanged over a radial range
from $\sim 10$ kpc out to 40 kpc  \citep{h99,h00,rej05};
it is only within the bulge region at $r < 8$ kpc (for NGC 5128,
$r_{eff} = 5.5$ kpc) that significant
changes began showing up and the MDF became systematically more
metal-rich (HH02).

Finally, we note that in either logarithmic or linear form,
the MDF is very broad but contains a tantalizing hint
of fine structure.  The ``notches'' at $(Z/Z_{\odot}) \simeq 0.2$ 
and at $\simeq$0.3 are statistically significant at $\gtrsim 5 \sigma$ in a nominal
sense (the bin at the first gap has 1000 stars whereas the smooth model would
predict 1500; and the second gap has 820 stars versus 1120 expected from the model
curve).  Furthermore, the same gaps appear in all three of the radial bins plotted in 
Fig.~\ref{z_3panel}.
These small gaps are suggestive, perhaps, of distinct episodes
of star formation that were incompletely averaged out over the 
whole galaxy.  The extreme
low-luminosity systems $\omega$ Cen and Carina show MDFs that are
more sharply divided into a small number of distinct episodes 
\citep[e.g.][among many others]{koch06,sollima05}, whereas no such gaps 
appear in the much larger NGC 5128 \citep{h02,rej05}.  In this respect
as in many others, NGC 3377 appears to occupy an intermediate position
in the whole range of old, composite stellar systems:  the smaller the
system, the more likely it is that its MDF will be dominated by a
smaller number of star-forming events and will take on a discontinuous
morphology.

\subsection{Comparisons with Smaller and Larger Systems}

Adding this study to some of the previously cited papers, we can now
make a first, admittedly rudimentary, attempt to put together a comparison
of MDFs for the full range of normal E galaxies, from dwarfs to giants.
An illustrative sample is shown in Figure \ref{z_4gal}.  To represent
the low-luminosity end of the normal E sequence, we use the ``typical''
Local Group dwarf NGC 147 ($M_V^T = -15.6$), with
photometric $(V,I)$ RGB data from \citet{han97}.  We also include
two spatial regions from the giant NGC 5128, at $M_V^T = -22.1$:  an outer-halo field
at 40 kpc projected distance \citep{rej05}, and the 8-kiloparsec
``inner halo'' or outer-bulge field analyzed by HH02.  In all cases, exactly the
same interpolation code and RGB model grid has been used to derive
the MDF, so the results are internally homogeneous.  

The sequence in Fig.~\ref{z_4gal}
runs from the small potential well of the dwarf, through the intermediate-mass
NGC 3377, to the outer regions of a giant, and finally to the deep inner
regions of the giant.  The change in the MDF clearly shows up as a progressively increasing
spread in heavy-element abundance.  We repeat that the upper tail for NGC 3377
is cut off by photometric incompleteness for $Z > 0.6 Z_{\odot}$ and it is quite
possible that a thin extension to higher $Z$ is present.  By contrast, the
NGC 147 and NGC 5128 color-magnitude diagrams are more deeply sampled and
less affected by incompleteness.   In terms of physical trends, however,
it is interesting to note that
the {\sl peak} of the MDF changes only modestly, from $\simeq 0.1 Z_{\odot}$
in the dwarf up to $\simeq 0.3 Z_{\odot}$ in the bulge of the giant.
Instead, the most prominent effect shows up in the relative number of stars at
high metallicity:  in the dwarf, nothing is present above $0.4 Z_{\odot}$,
while the MDF for the giant extends far above Solar.  This is the most
dramatic effect of the depth of the potential well within which their stars
were forming:  in the deeper wells, almost all of the gas will be retained,
permitting the enrichment to continue on through many generations and
go up to high $Z$.  A secondary feature worth noting is the steepness
of the rise at low $Z$ in the two smaller galaxies, showing the importance
of infalling gas.  This pattern is suggestive of an interpretation
that the smaller galaxies took somewhat longer to assemble than the giants,
during which time the inflow of pristine gas could continue longer
and have a more important effect on the final MDF.

\section{Summary}

We have used deep HST/ACS images of a field in the intermediate-luminosity
elliptical galaxy NGC 3377, a member of the Leo group, to derive the 
metallicity distribution function of its stars.  More than 57000 stars were
measured over a radial region extending from $1\farcm6$ to $5\farcm7$ from
galaxy center.  The $(V,I)$ data reach deep enough to reveal almost 2 magnitudes
of the red-giant branch and show the expected large spread in color that
accompanies a high internal dispersion in metallicity.  No significant presence
of any young stellar population (stars younger than a few Gyr) can be seen in
the color-magnitude diagram and the interpretation of the system is relatively
simple.

Using a finely spaced grid of RGB models calibrated against the Milky Way
globular clusters, we derive the heavy-element abundance distribution $n(Z)$
for the NGC 3377 stars.  No apparent radial gradient in metallicity shows up;
in all parts of the halo, the distribution peaks at log $(Z/Z_{\odot}) \simeq -0.6$
and is quite broad.  Very few stars are, however, more metal-poor than
log $(Z/Z_{\odot}) = -1.5$, and the data suggest that few stars are
more metal-rich than $-0.3$.  It is possible that
this galaxy did not reach Solar abundance during its enrichment history, 
at least in its halo.  The overall shape
of its MDF in all respects lies between the more metal-poor dwarf ellipticals
on the one hand, and the very much more metal-rich giants on the other.

\acknowledgements

WEH and GLHH thank the Natural Sciences and Engineering Research Council of Canada for
financial support.  We also thank an anonymous referee for constructive
comments on the first version of the paper.

\clearpage

\begin{deluxetable}{ccccccccc}
\tablecaption{Basic Parameters for NGC 3377 \label{basics}}
\tablewidth{0pt}
\tablehead{
\colhead{Parameter} &  \colhead{Value} \\
}

\startdata
$Type$ & E5 \\
$\alpha$ (J2000) & $10^h 47^m 42\fs4$ \\
$\delta$ (J2000) & $13\fdg 59' 08''$ \\
$v_r$ (helio) & 665 km s$^{-1}$ \\
$A_V$ & 0.10 \\
$(m-M)_0$ & $30.17 \pm 0.10$ \\
$V_T^0$ & 10.23 \\
$M_V^T$ & $-19.9$ \\

\enddata

\end{deluxetable}

\clearpage

%
%
\begin{figure} 
\figurenum{1} 
\plotone{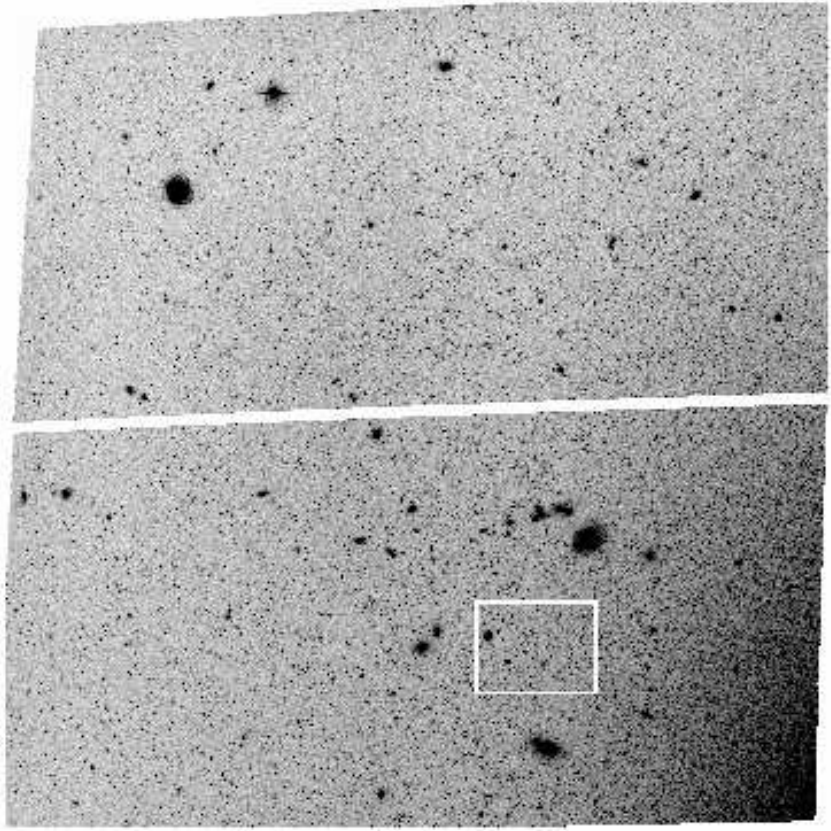} 
\caption{A reproduction of our measured ACS field in NGC 3377.  Reproduced here
is the combined $F814W$ ($I$ band) image.  At the $\sim 11$-Mpc distance
of NGC 3377, the ACS/WFC field size corresponds to a width of 11 kpc.
The inset box shows the location of
the smaller field reproduced in the next Figure.  Note that the center of the
galaxy is well off the field to the lower right; the ACS field center is $3\farcm8$
away from the galaxy center.}
\label{widefield}
\end{figure}

\begin{figure} 
\figurenum{2} 
\plotone{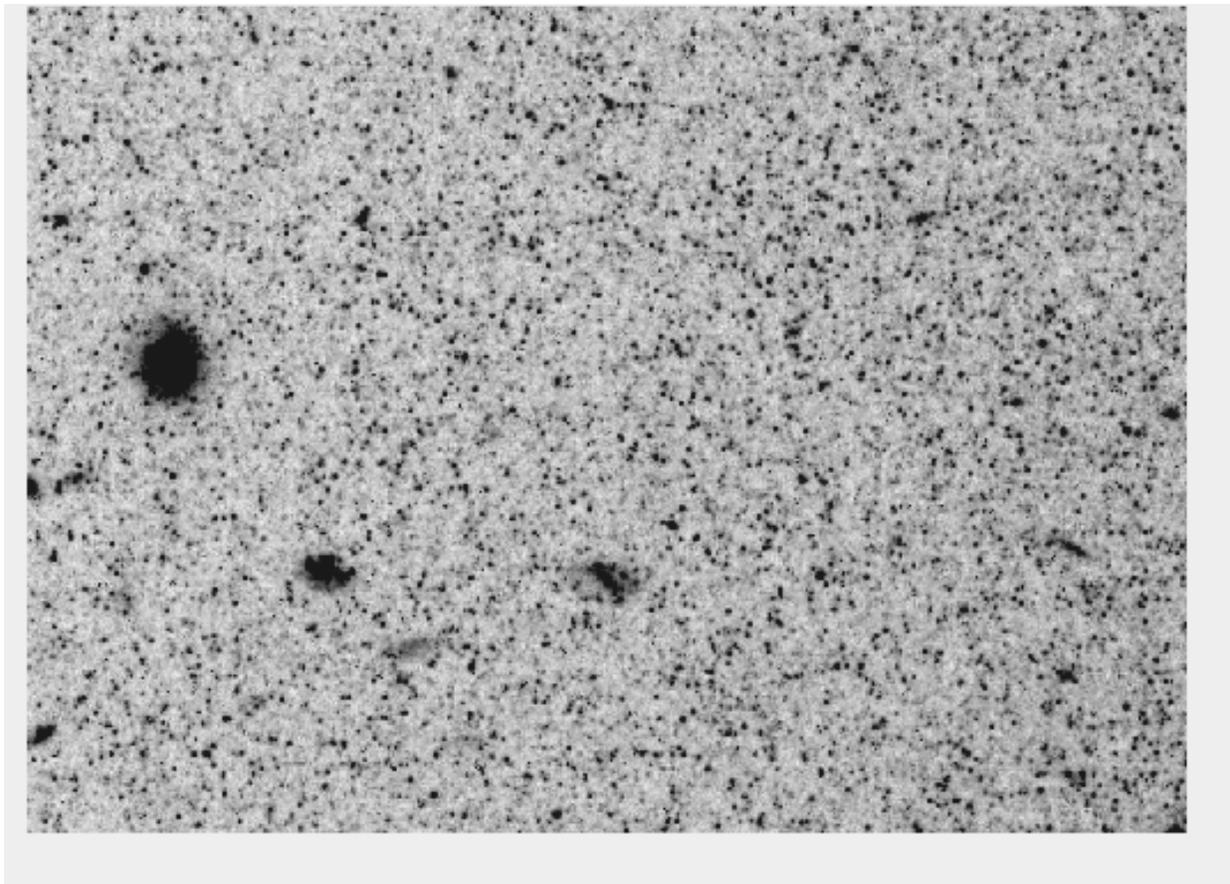} 
\caption{A small section of the NGC 3377 ACS field.  The image shown here is from the
composite $I-$band exposure and is $36''$ (700 pixels) wide.  The halo red-giant stars
are the small starlike points scattered everywhere across the field; a few distant
background galaxies are also visible.}
\label{smallfield}
\end{figure}

\begin{figure} 
\figurenum{3} 
\plotone{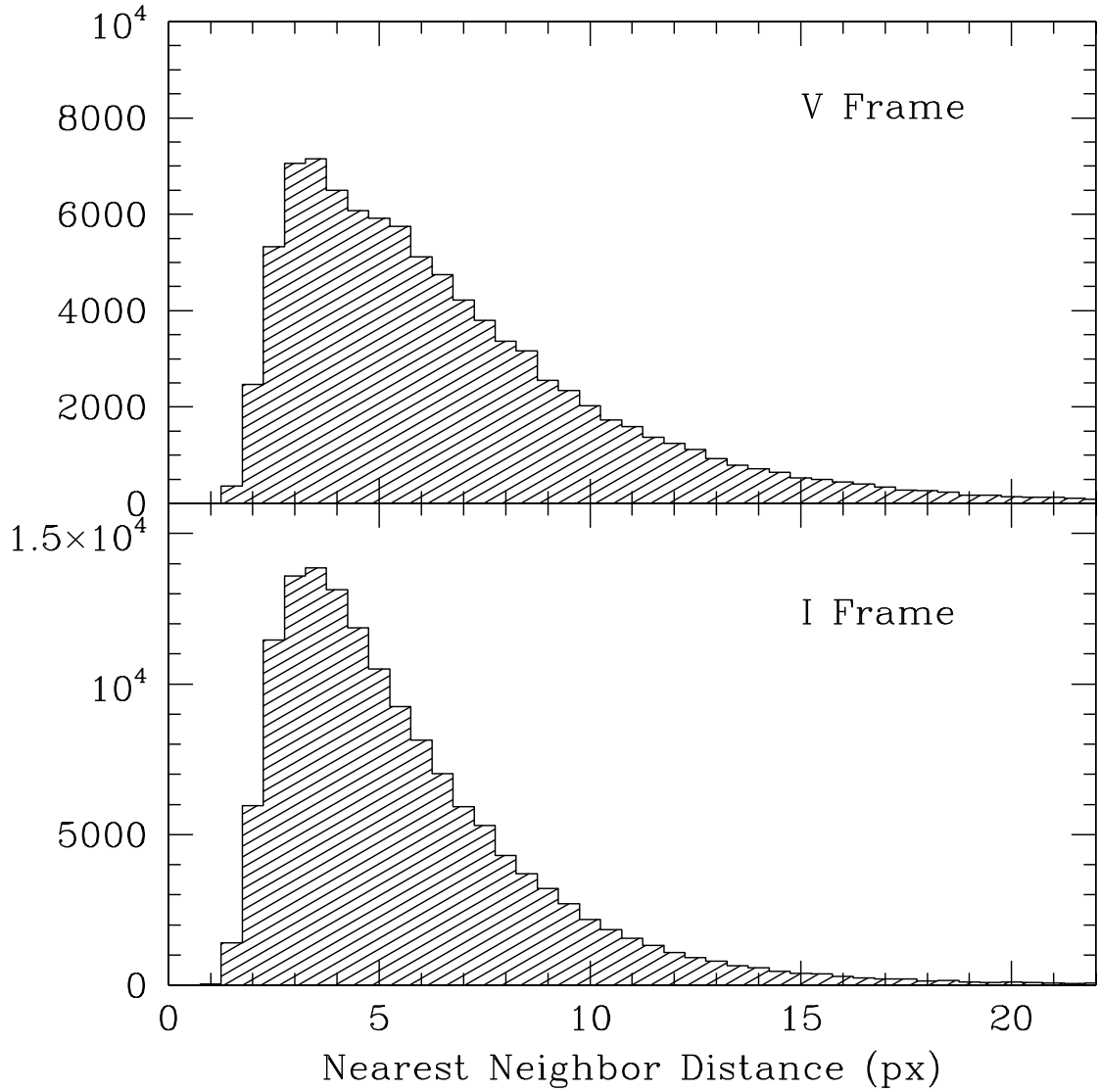} 
\caption{Histograms of the nearest-neighbor distances (NND) for all detected
objects on each of the two frames.  We reject any objects with $NND < 3$ px
from the subsequent analysis to minimize any effects of crowding (see text)
}
\label{crowding}
\end{figure}

\begin{figure} 
\figurenum{4} 
\plotone{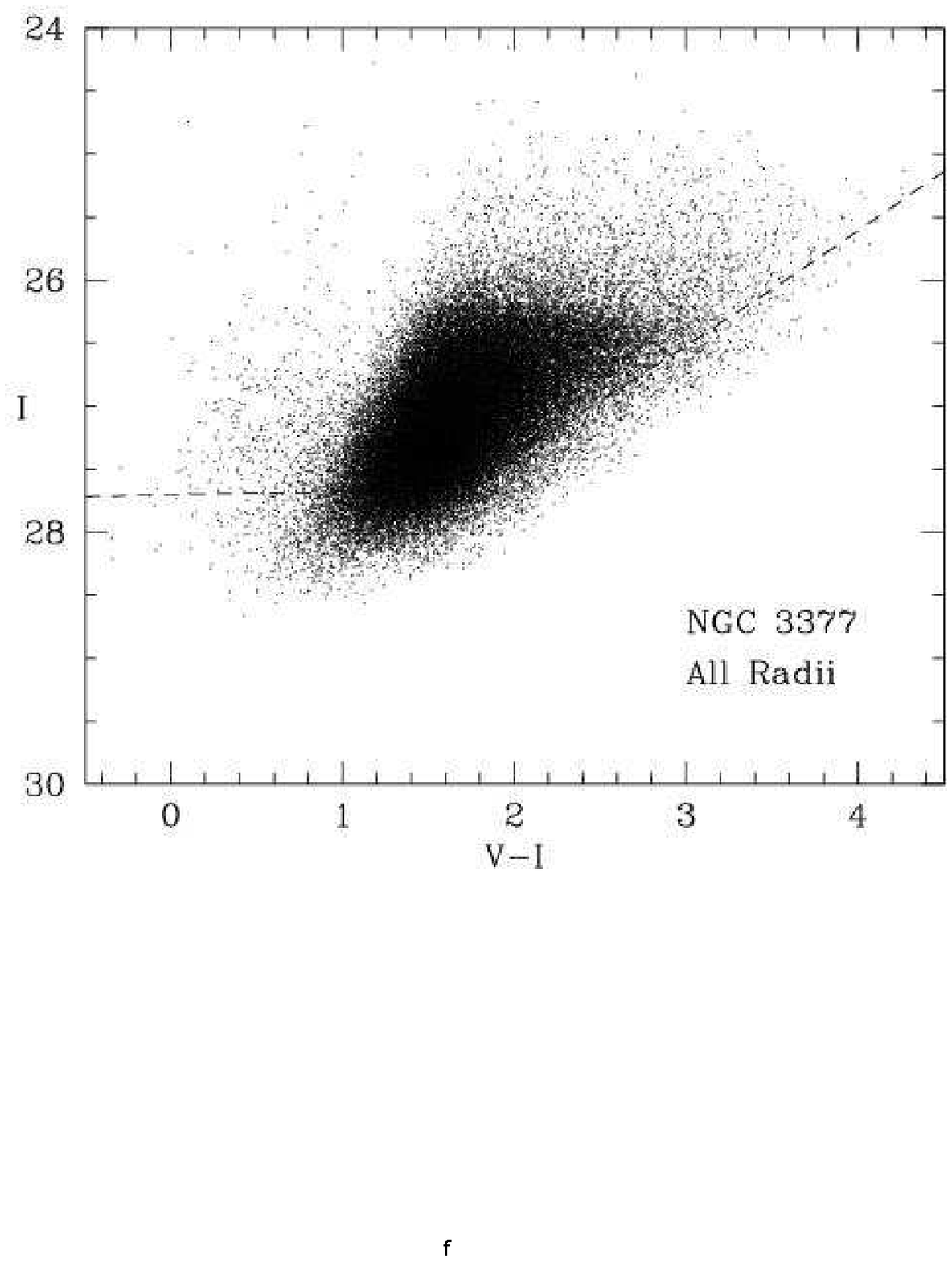} 
\caption{The color-magnitude array in $(V,I)$ for our sample of halo stars in NGC 3377.
All 70,647 measured stars with high quality photometry are shown.  The dashed lines show
the 50\% completeness limits of our measurements, as derived from artificial-star
tests (see text).}
\label{cmd1}
\end{figure}

\begin{figure} 
\figurenum{5} 
\plotone{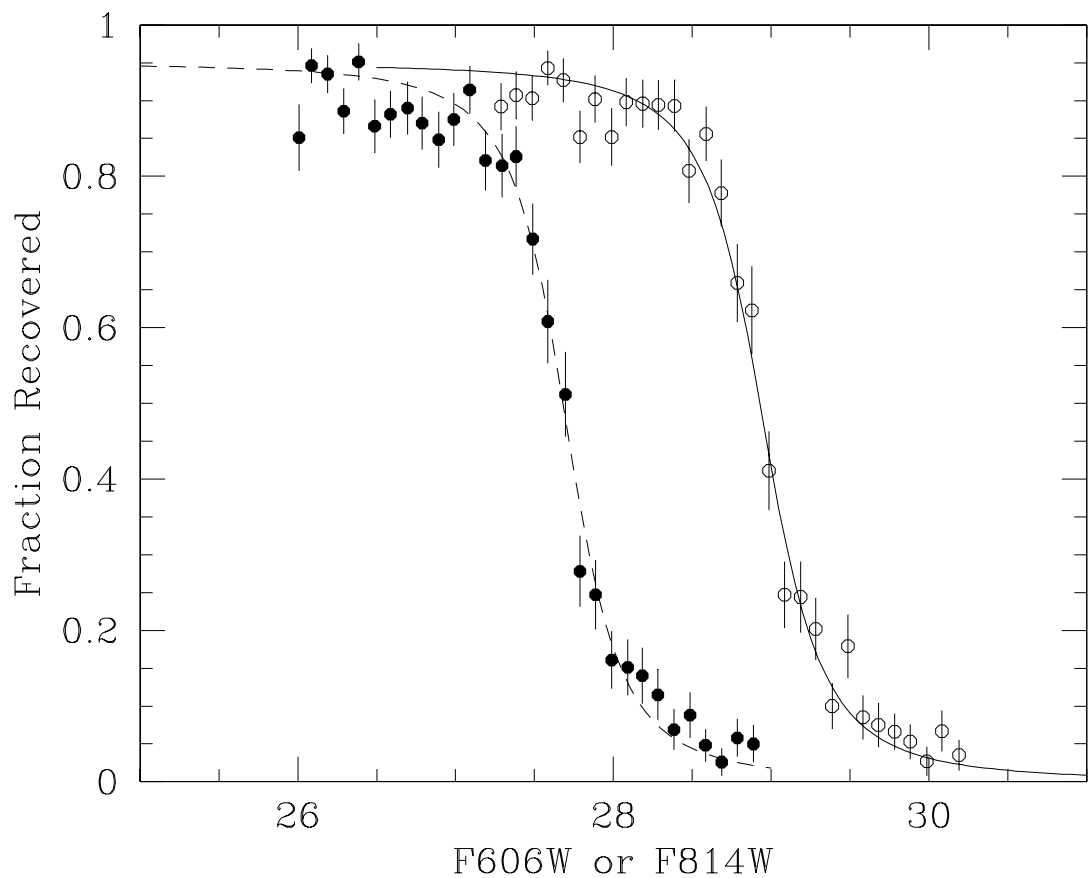} 
\caption{Photometric detection completeness functions, as evaluated by
artificial-star experiments.  The vertical axis plots the
number of stars recovered by our {\sl DAOPHOT find/allstar} measurement sequence
divided by the number artificially inserted at the magnitude given along the
horizontal axis.  The {\sl solid symbols} are the $F814W$ ($I-$band) stars and
the {\sl open symbols} are for $F606W$ ($V-$band). The fitted curves drawn through
each set of points are Pritchet functions with parameters as given in the text.} 
\label{completeness} 
\end{figure}

\begin{figure}
\figurenum{6}
\plotone{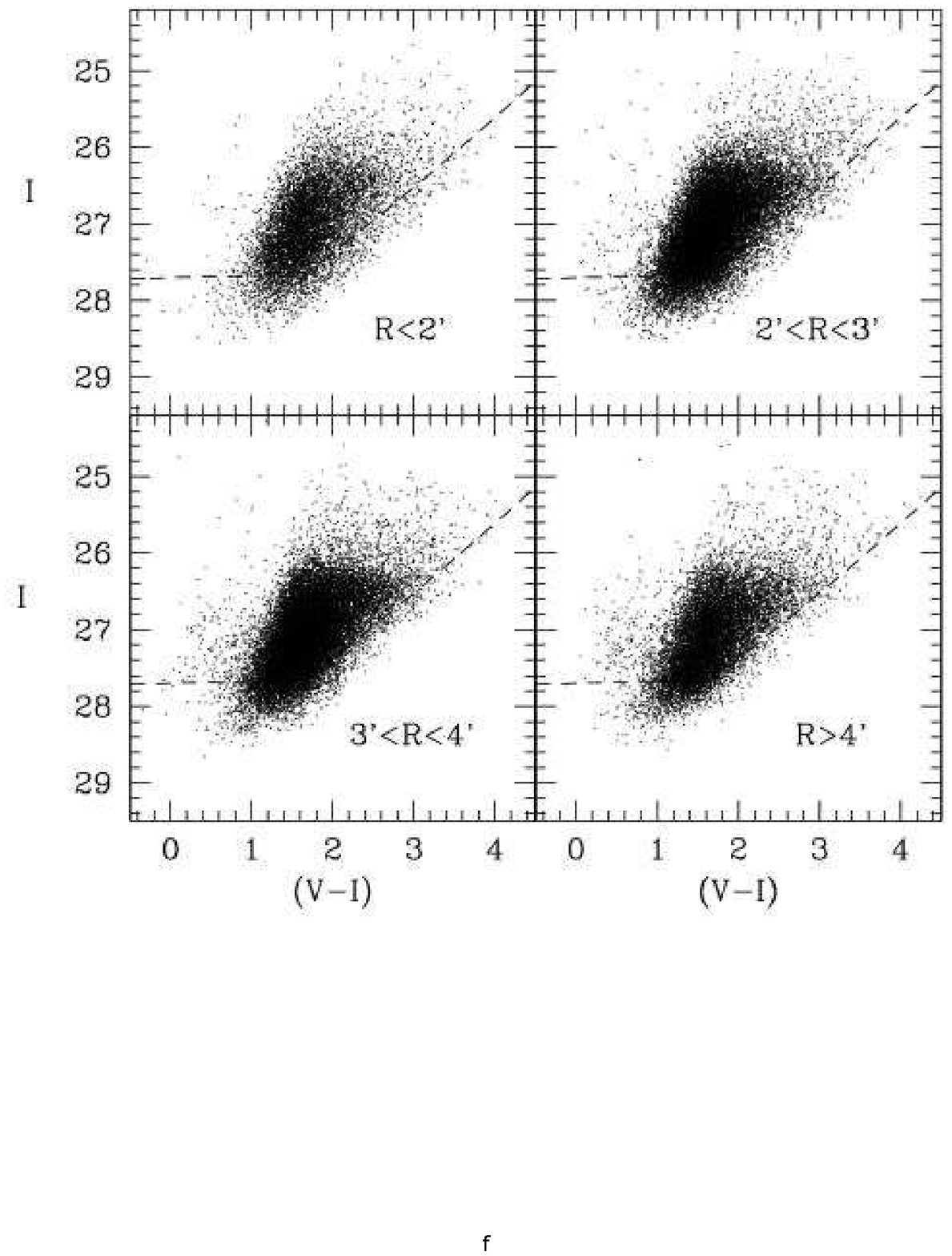}
\caption{Color-magnitude diagrams subdivided by region.  Here $R$
is the distance from the center of NGC 3377 in arcminutes.}
\label{4panel_cmd}
\end{figure}

\begin{figure}
\figurenum{7}
\plotone{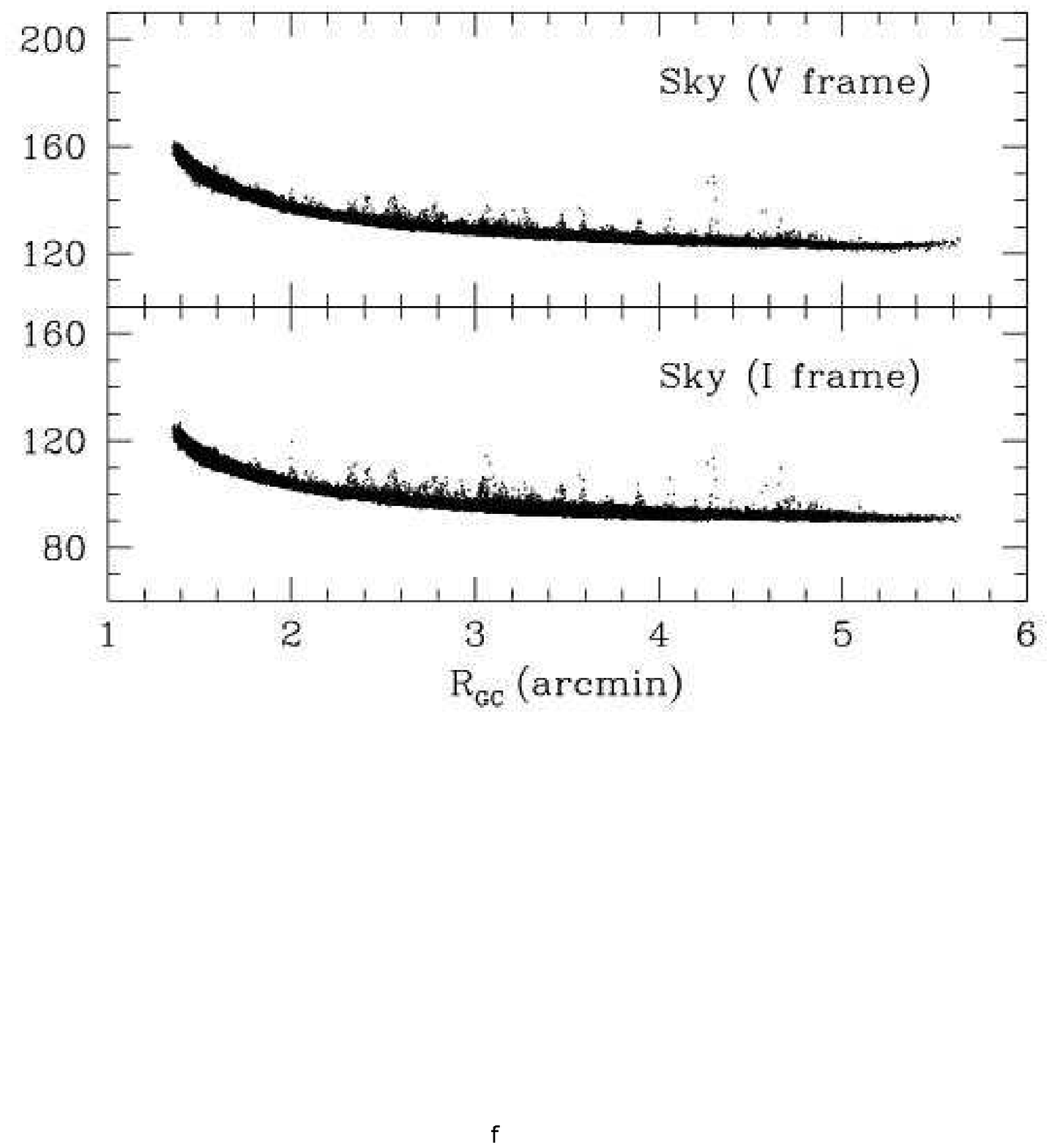}
\caption{Plot of the local sky intensity level as a function of
radius from the center of NGC 3377.  Here the vertical scale is
the local sky level around each star, in units of
DU as returned by {\sl allstar}.  The vertical normalization is
arbitrary.}
\label{skylevel}
\end{figure}

\begin{figure}
\figurenum{8}
\plotone{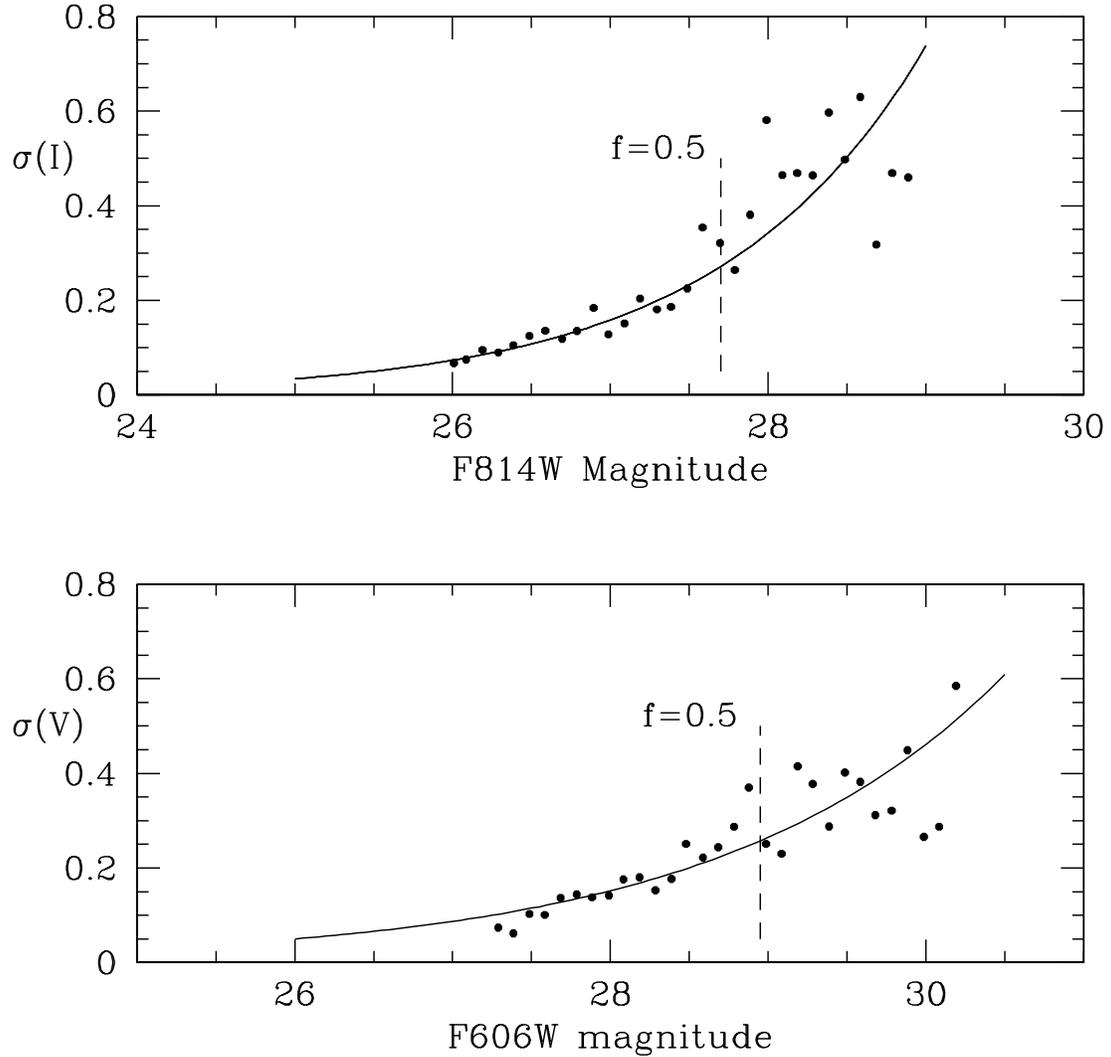}
\caption{Random measurement uncertainties of the photometry, as derived
from the artificial-star tests.  Each plotted point (in 0.02-magnitude bins)
gives the root-mean-square
scatter between the input and measured magnitude for all stars in the bin.
The mean lines are exponential curves with parameters as given in the text.
The two vertical dashed lines show the limiting magnitudes of the photometry,
i.e.~the magnitude levels at which the detection completeness is 50\%.}
\label{random_errors}
\end{figure}

\begin{figure} 
\figurenum{9} 
\plotone{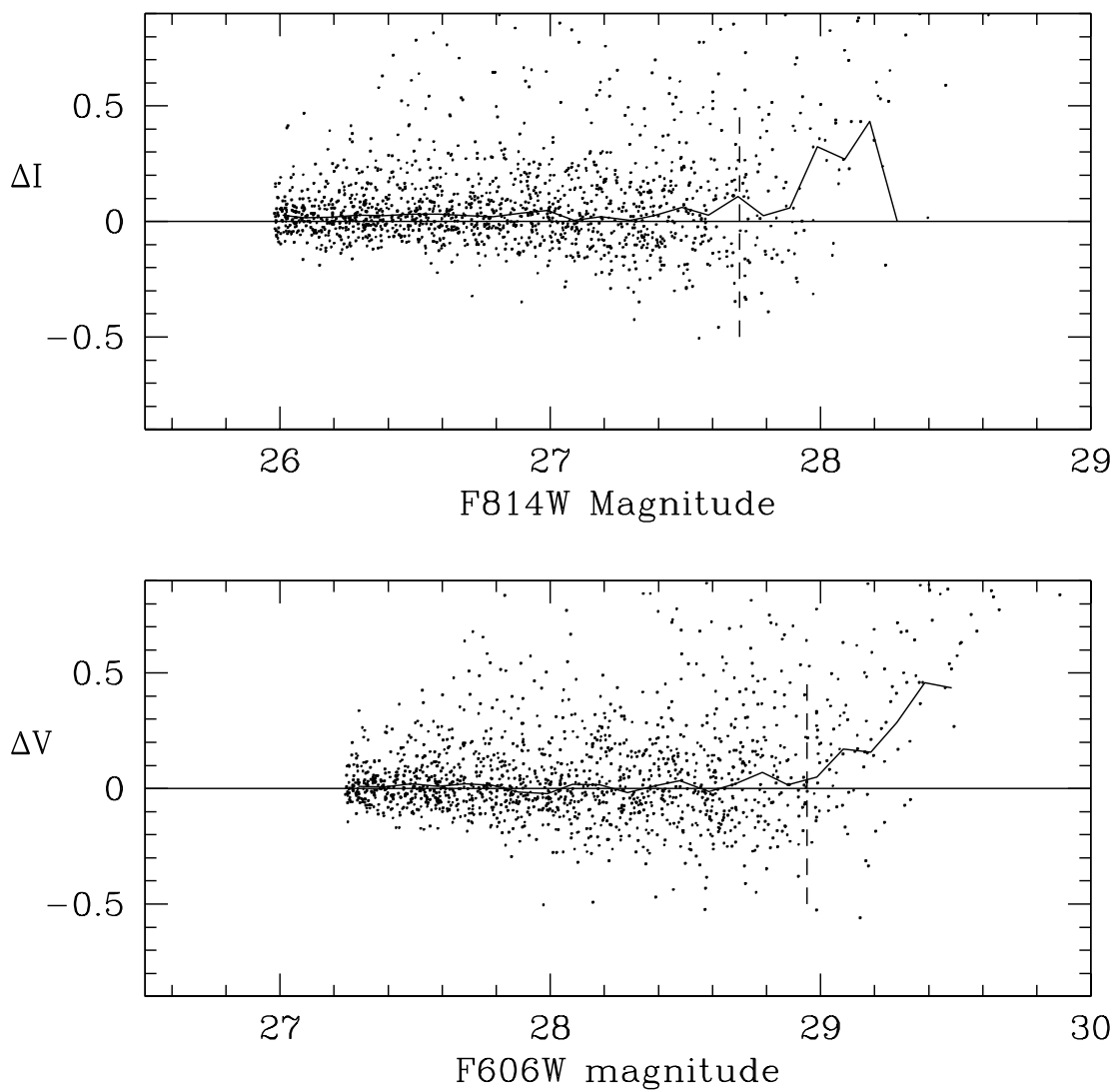} 
\caption{Random and systematic errors of the photometry, as derived from
artificial-star tests.  Each graph shows the difference between the inserted
and measured magnitudes of the fake stars in the sense (input minus measured),
so that stars measured ``too bright'' have positive values in the plot.
The solid line through each plot at 0.1-magnitude bins is the median $\Delta m$
within each bin.  The vertical {\sl dashed line} in each plot shows the
``completeness limit'' (the magnitude, from the previous figure, at which
the detection completeness drops to 50 percent). Systematic errors increase
dramatically for stars fainter than the completeness limit.}
\label{photom_errors}
\end{figure}

\begin{figure} 
\figurenum{10} 
\plotone{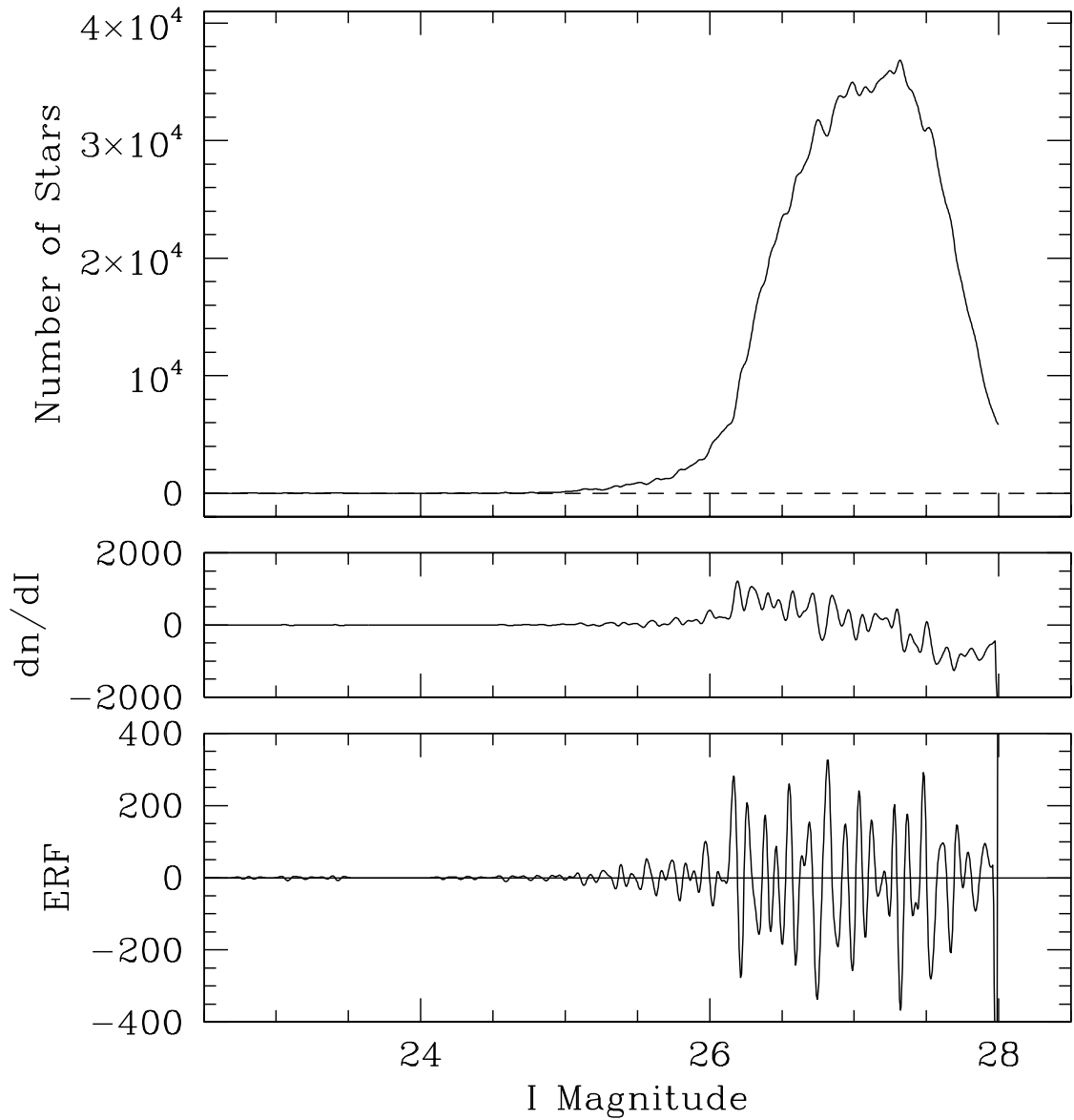} 
\caption{Luminosity function for all measured stars in the NGC 3377 field.
In the uppermost graph, the number of stars per unit magnitude is plotted 
against $I$; the faint-end steep decline is driven by strong photometric
incompleteness.  The middle panel shows our numerical estimate for
the first derivative $dn/dI$ of the LF, while the lower panel shows the numerical
estimate for the second derivative $d^2n/dI^2$. In all panels the data have
been convolved with a Gaussian smoothing kernel of $\sigma = 0.02$ mag.}
\label{lf} 
\end{figure}

\begin{figure} 
\figurenum{11} 
\plotone{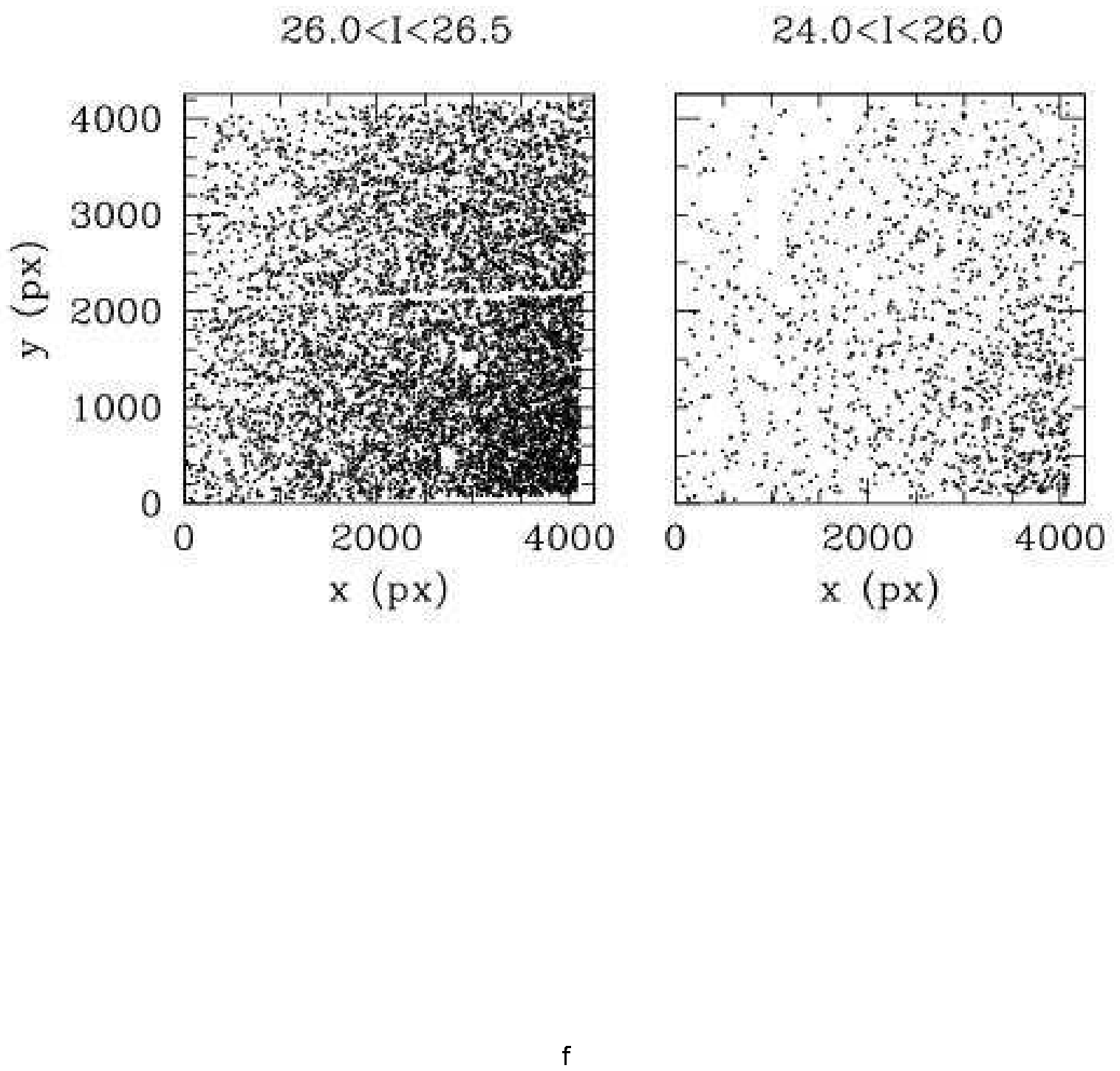} 
\caption{{\sl Left panel:} Locations of stars in the brightest half-magnitude
of the NGC 3377 red-giant branch.  The location of the gap between the two
ACS/WFC chips can be seen through the middle, as well as small
cutout masked regions around some background galaxies.
{\sl Right panel:}  Locations of stars brighter than the RGB tip.}
\label{xy2} 
\end{figure}

\begin{figure} 
\figurenum{12} 
\plotone{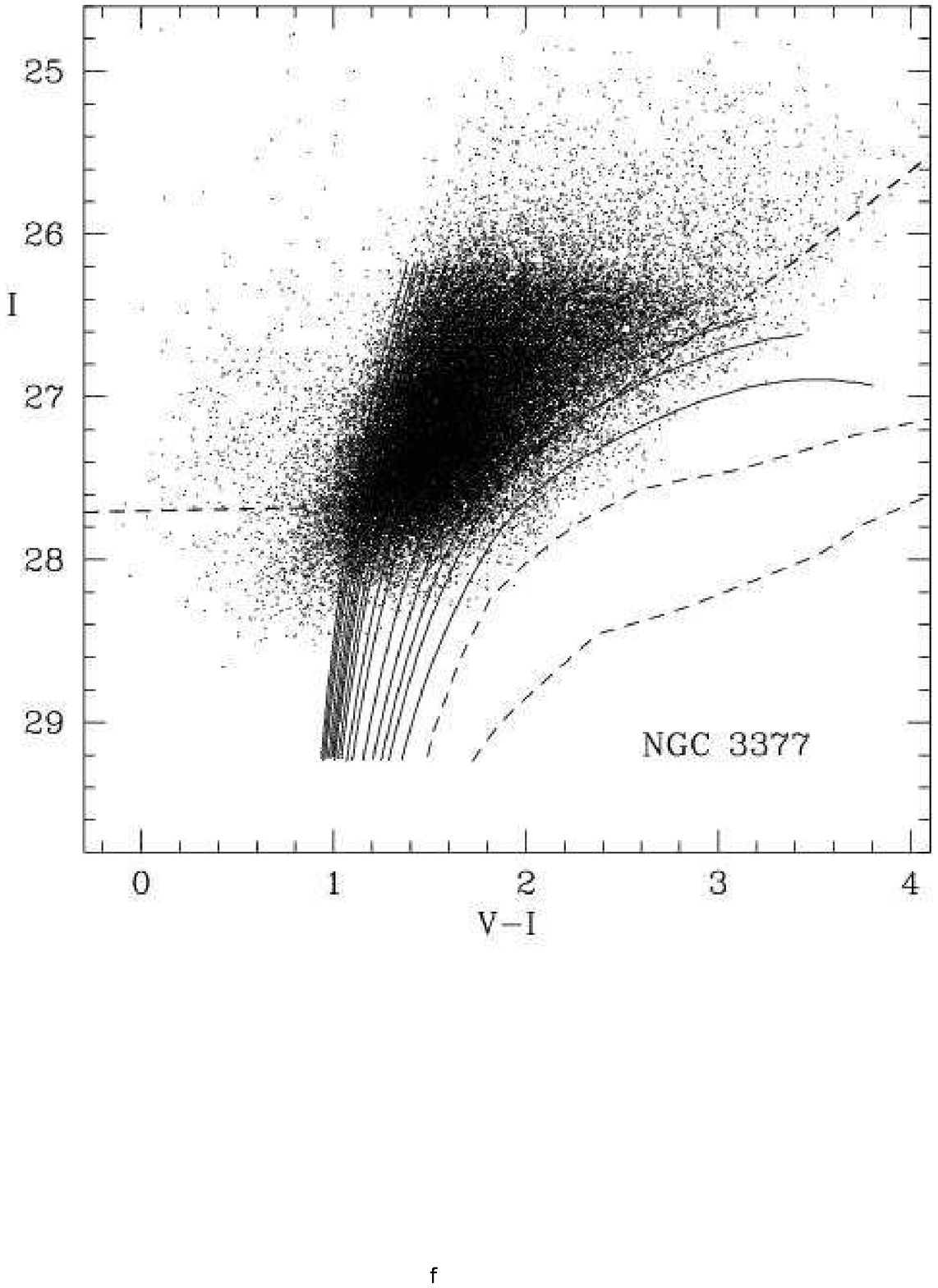} 
\caption{Color-magnitude diagram for the NGC 3377 red giants, with model
red-giant tracks superimposed on the data points.  All tracks are for
ages of 12 Gy, but differ in metallicity roughly in steps of 
$\Delta$[Fe/H] $\simeq 0.1$.  These are the same set of tracks from
VandenBerg et al.~(2000) we have used for our previous study of the
NGC 5128 halo, supplemented by two 
metal-rich tracks generated from old Milky Way star clusters
(see Harris \& Harris 2002), extending from log $(Z/Z_{\odot}) = -2.0$
to $+0.4$.  The placement of the
tracks assumes a foreground reddening $E(V-I) = 0.02$ and a
distance modulus $(m-M)_I = 30.25$ as derived in the text.
The 50\% photometric completeness limit is shown as the pair
of dashed lines.}
\label{cmd_fiducial} 
\end{figure}

\begin{figure} 
\figurenum{13} 
\plotone{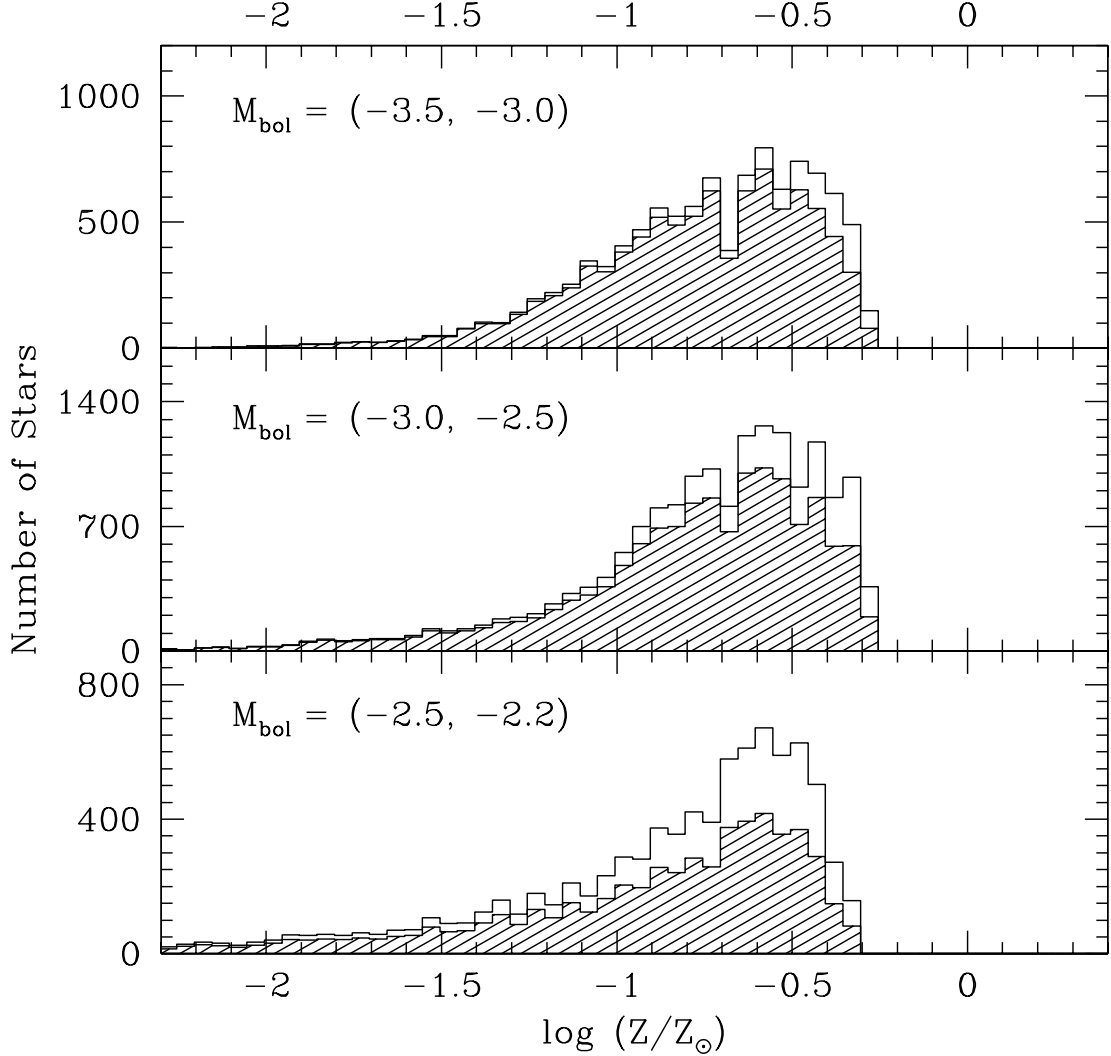} 
\caption{Metallicity distribution function for the halo red giants, divided
into three approximate luminosity bins.  This division tests for any systematic
errors in the interpolation routine or the placement of the evolutionary tracks.
The {\sl shaded regions} show the MDF uncorrected for photometric completeness,
while the higher {\sl unshaded regions} show the full completeness-corrected MDF.
Stars more metal-rich than log $(Z/Z_{\odot}) \simeq -0.3$ fall to the red
of the 50\% photometric completeness line (see previous Figure) and are not
included in the sample.}
\label{feh_3panel} 
\end{figure}

\begin{figure} 
\figurenum{14} 
\plotone{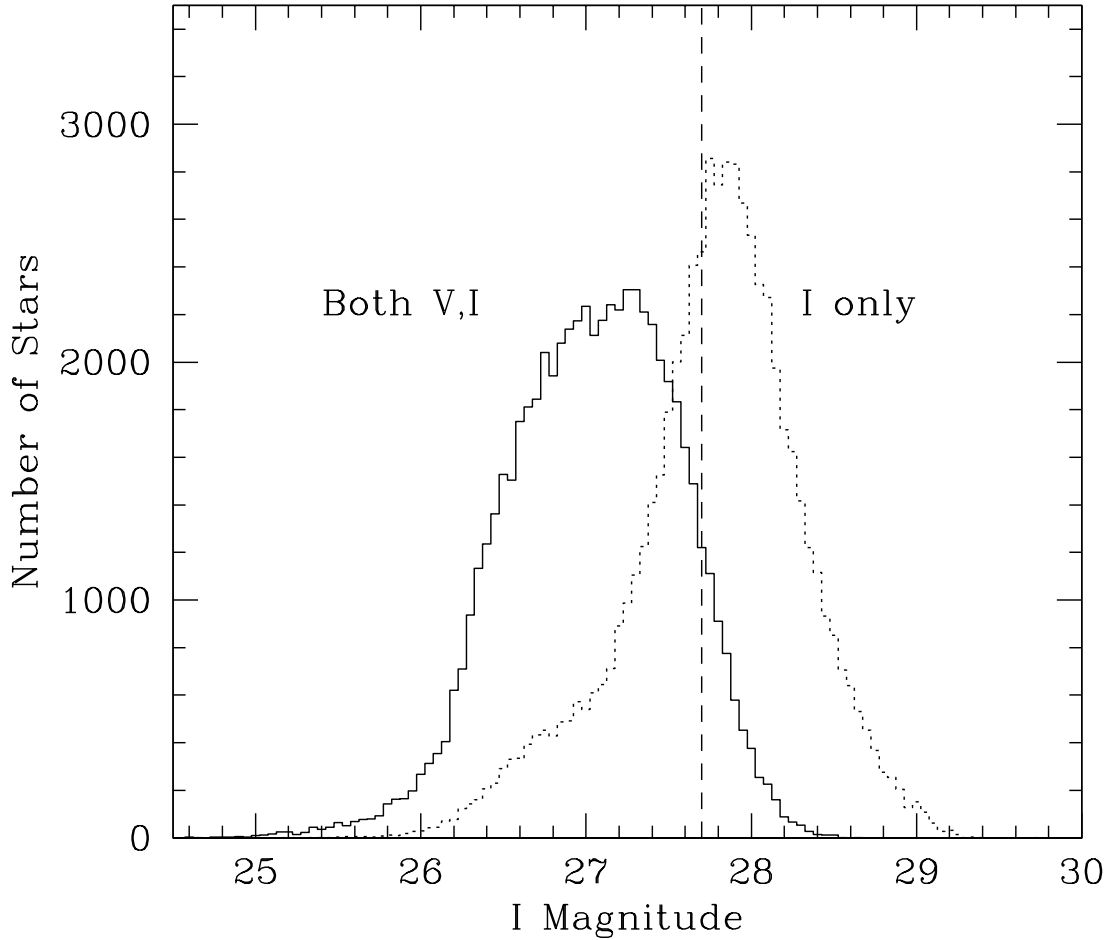} 
\caption{Luminosity function for the NGC 3377 stars.  The vertical axis
here is rescaled from Figure 10 to show the number of stars
per 0.05-mag bin plotted against $I$ magnitude.  The {\sl solid line}
shows the data for the stars in the previous figure, 
with measurements in both $V$ and $I$.  The {\sl dotted line}
shows stars measured {\sl only} in $I$ but not in $V$, thus ones that
do not appear in the color-magnitude diagram.
The vertical dashed line at $I = 27.7$ is the 50\% photometric completeness
level in $I$.  Note that for red stars measured in both $V$ and $I$ the 
detection completeness is severely incomplete fainter than $V \simeq 26.6$,
thus the turnover in the solid curve is driven by the $V$ filter cutoff.}
\label{2lf} 
\end{figure}

\begin{figure} 
\figurenum{15} 
\plotone{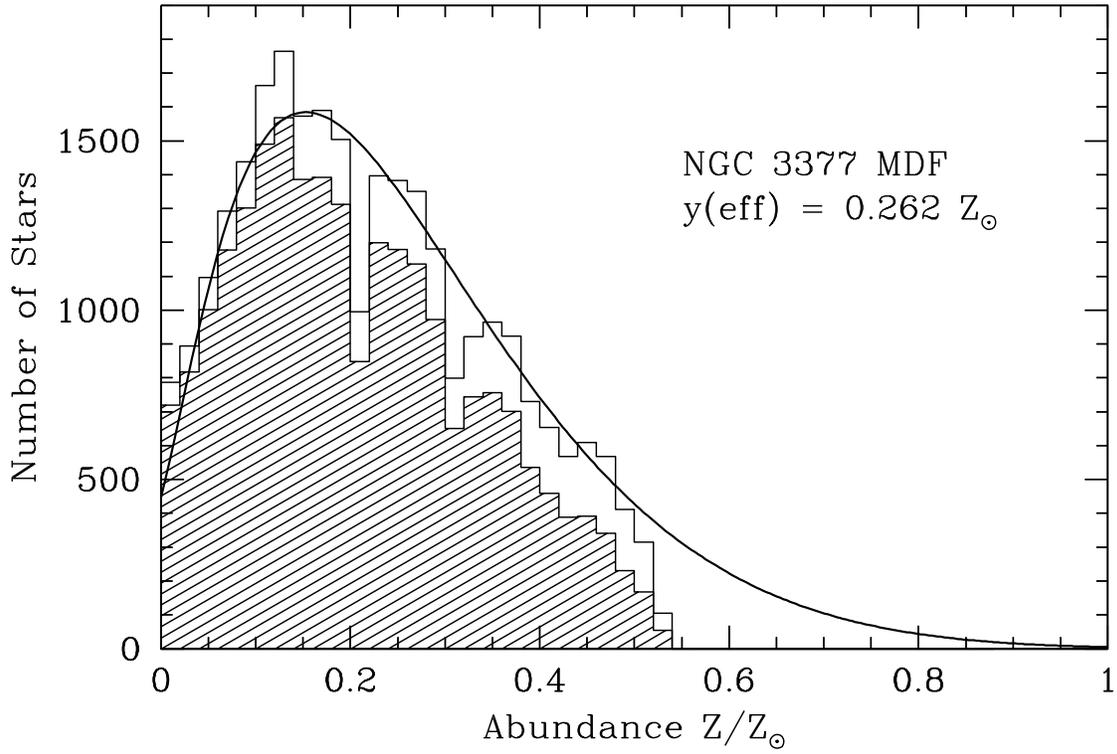} 
\caption{Heavy-element abundance distribution, plotted in linear form.
The number of stars per unit bin $\Delta Z = 0.02$ is plotted against
$Z/Z_{\odot}$.  The solid curve is a chemical evolution model with exponentially
declining infall, as discussed in the text.  The assumed effective yield 
(the combination of stellar nucleosynthesis plus outflow; see text) 
is $y_{eff} = 0.0039 \simeq 0.26 Z_{\odot}$.}
\label{zmodel} 
\end{figure}

\begin{figure} 
\figurenum{16} 
\plotone{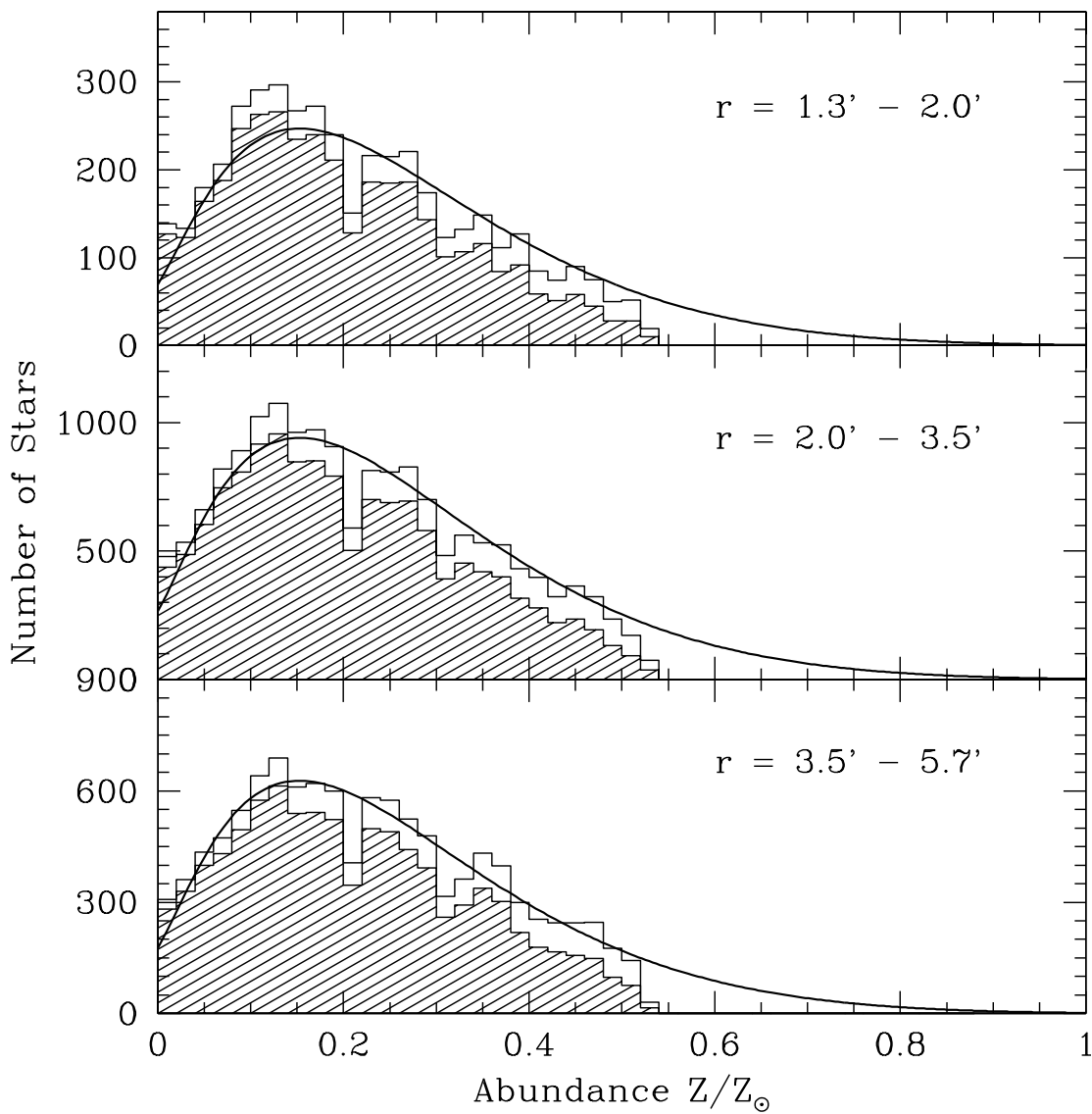} 
\caption{Heavy-element abundance distribution for three different
radial ranges.  The projected radial range in arcminutes is shown in 
each panel, from innermost to outermost.  As in the previous Figure, the
shaded region shows the raw number of stars in each $Z-$bin and the
unshaded region the completeness-corrected bin totals.  The solid curve
in each case is the same chemical evolution model that was applied
to the total population in the previous Figure.  All three regions
shown are outside the effective radius $R_e = 1\farcm1$ of the galaxy.
}
\label{z_3panel} 
\end{figure}

\begin{figure} 
\figurenum{17} 
\plotone{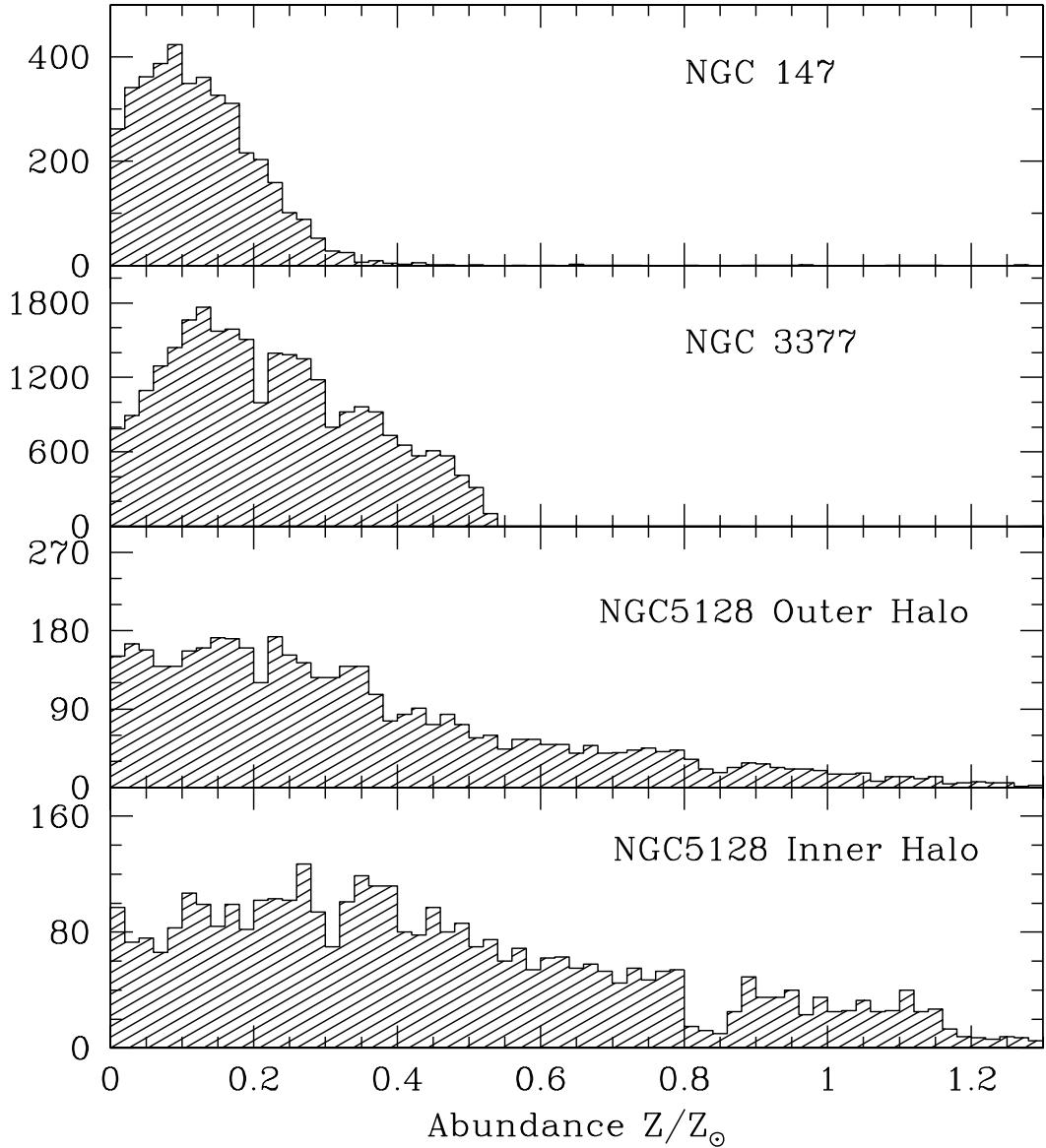} 
\caption{Heavy-element abundance distributions for four different samples
of stars. From top to bottom these are: (a) the Local Group dwarf elliptical
NGC 147, with data from Han et al. (1997); (b) the Leo elliptical NGC 3377
(this study); (c) the outer halo of the giant elliptical NGC 5128, from \protect\cite{rej05};
and (d) the inner halo of NGC 5128, from \protect\cite{h02}.
}
\label{z_4gal} 
\end{figure}

\begin{thebibliography}{}

\bibitem[Bahcall \& Soneira(1981)]{bah81} Bahcall, J., \& Soneira, R. 1981, 
\apjs, 47, 357  
\bibitem[Beasley et al.(2003)]{bea03} Beasley, M.A., Harris, W.E., Harris, G.L.H., 
\& Forbes, D.A. 2003, \mnras, 340, 341
\bibitem[Bekki et al.(2003)]{bekki03} Bekki, K., Harris, W.E., \& Harris, G.L.H.
2003, \mnras, 338, 587
\bibitem[Bellazzini et al.(2004)]{bell04} Bellazzini, M., Ferraro, F.R., 
Sollima, A., Pancino, E., \& Origlia, L. 2004, \aap, 424, 199
\bibitem[Binney \& Merrifield(1998)]{binney98} Binney, J., \& Merrifield, M. 1998, 
Galactic Astronomy (Princeton: Princeton Univ. Press)
\bibitem[Chiappini et al.(1997)]{chia97} Chiappini, C., Matteuci, F., \& Gratton, R. 
1997, \apj, 477, 765
\bibitem[Ciardullo et al.(1989)]{ciar89} Ciardullo, R., Jacoby, G.H.,
\& Ford, H.C. 1989, \apj, 344, 715
\bibitem[Ciardullo et al.(2002)]{ciar02} Ciardullo, R., Feldmeier, J.J., Jacoby, G.H.,
Kuzio de Naray, R., Laychak, M.B., \& Durrell, P.R. 2002, \apj, 577, 31
\bibitem[Cole et al.(2000)]{cole00} Cole, S., Lacey, C.G., Baugh, C.M., \& 
Frenk, C.S. 2000, \mnras, 319, 168
\bibitem[Fleming et al.(1995)]{fl95} Fleming, D.E.B., Harris, W.E., 
Pritchet, C.J., \& Hanes, D.A. 1995, \aj, 109, 1044
\bibitem[Freedman et al.(2001)]{freedman01} Freedman, W.L. et al. 2001,
\apj, 553, 47
\bibitem[Graham et al.(1997)]{graham97} Graham, J.A. et al. 1997, \apj, 477, 535
\bibitem[Gregg et al.(2004)]{gregg04} Gregg, M.D., Ferguson, H.C., Minniti, D.,
Tanvir, N., \& Catchpole, R. 2004, \aj, 127, 1441
\bibitem[Grillmair et al.(1996)]{gri96} Grillm,air, C.J. et al. 1996, \aj, 112, 1975
\bibitem[Han et al.(1997)]{han97} Han, M. et al. 1997, AJ, 113, 1001
\bibitem[Harris, Harris, \& Poole(1999)]{h99} Harris, G.L.H., Harris, W.E., \& 
Poole, G.B. 1999, \aj, 117, 855
\bibitem[Harris \& Harris(2000)]{h00} Harris, G.L.H., \& Harris, W.E. 2000, 
\aj, 120, 2423
\bibitem[Harris(1996)]{harris96} Harris, W.E. 1996, \aj, 112, 1487
\bibitem[Harris(2001)]{har01} Harris, W.E. 2001, in Star Clusters,
Saas-Fee Advanced Course 28 (New York: Springer), ed.~L.Labhardt \& B.Binggeli
\bibitem[Harris et al.(1998)]{har98} Harris, W.E., Durrell, P.R., Pierce, M.J.,
\& Secker, J. 1998, Nature, 395, 45
\bibitem[Harris \& Harris(2002)]{h02} Harris, W.E., \& Harris, G.L.H. 2002, \aj, 
123, 3108 (HH02)
\bibitem[Kawata \& Gibson(2003)]{kawata03} Kawata, D., \& Gibson, B.K. 2003,
\mnras, 340, 908
\bibitem[Koch et al.(2006)]{koch06} Koch, A., Grebel, E.K., Wyse, R.F.G., 
Kleyna, J.T., Wilkinson, M.I., Harbeck, D.R., Gilmore, G.F., \& Evans, N.W. 
2006, \aj, 131, 895
\bibitem[Lafranchi \& Matteuci(2004)]{laf04} Lafranchi, G.A., \& Matteucci, F.
2004, \mnras, 351, 1338
\bibitem[Landolt(1992)]{lan92} Landolt, A.U. 1992, \aj, 104, 340
\bibitem[Naab \& Ostriker(2006)]{naab06} Naab, T., \& Ostriker, J.P. 2006, \mnras, 366, 899
\bibitem[Oey(2000)]{oey00} Oey, S. 2000, \apj, 542, L25
\bibitem[Pagel \& Patchett(1975)]{pagel75} Pagel, B.E.J., \& Patchett, B.E. 1975, 
\mnras, 172, 13
\bibitem[Prantzos(2003)]{prantzos03} Prantzos, N. 2003, \aap,404, 211
\bibitem[Rejkuba et al.(2003)]{rej03} Rejkuba, M., Minnite, D., 
Silva, D.R., \& Bedding, T.R. 2003, \aap, 411, 351
\bibitem[Rejkuba et al.(2005)]{rej05} Rejkuba, M., Greggio, L., 
Harris, W.E., Harris, G.L.H., \& Peng, E.W. 2005, \apj, 631, 262
\bibitem[Renzini(1998)]{ren98} Renzini, A. 1998, \aj, 115, 2459
\bibitem[Romano et al.(2005)]{rom05} Romano, D., Tosi, M., \& Matteucci, F. 2005,
\mnras, 365, 759
\bibitem[Saha et al.(1999)]{saha99} Saha, A., Sandage, A., Tamman, G.A., 
Labhardt, L., Maccheto, F.D., \& Panagia, N. 1999, \apj, 522, 802
\bibitem[Sakai et al.(1996)]{sakai96} Sakai, S., Madore, B.F., \& Freedman, W.L.
1996, \apj, 461, 713
\bibitem[Sakai et al.(1997)]{sakai97} Sakai, S., Madore, B.F., Freedman, W.L.,
Lauer, T.R., Ajhar, E.A., \& Baum, W.A. 1997, \apj, 478, 49
\bibitem[Salaris et al.(2002)]{sal02} Salaris, M., Cassisi, S., \& Weiss, A. 2002,
\pasp, 114, 375
\bibitem[Scannapieco et al.(2005)]{scan05} Scannapieco, C., Tissera, P.B., White, S.D.M.,
\& Springel, V. 2005, \mnras, 364, 552
\bibitem[Sirianni et al.(2005)]{sir05} Sirianni, M. et al. 2006, \pasp, 117, 1049
\bibitem[Sollima et al.(2005)]{sollima05} Sollima, A., Pancino, E., Ferraro, F.R., 
Bellazzini, M., Straniero, O., \& Pasquini, L. 2005, \apj, 634, 332
\bibitem[Somerville \& Primack(1999)]{som99} Somerville, R.S., \& Primack, J.R. 1999, 
\mnras, 310, 1087
\bibitem[Stetson(2000)]{ste00} Stetson, P.B. 2000, \pasp, 112, 925
\bibitem[Stetson(2005)]{ste05} Stetson, P.B. 2005, \pasp, 117, 563
\bibitem[Tanvir et al.(1999)]{tanvir99} Tanvir, N.R., Ferguson, H.C., \& Shanks, T.
1999, \mnras, 310, 175
\bibitem[Tonry et al.(2001)]{tonry01} Tonry, J.L., Dressler, A., Blakeslee, J.P., 
Ajhar, E.A., Fletcher, A.B., Luppino, G.A., Metzger, M.R., \& Moore, C.B. 2001,
\apj, 546, 681
\bibitem[Valle et al.(2005)]{valle05} Valle, G., Shore, S.N., \& Galli, D.
2005, \aap, 435, 551
\bibitem[Vandalfsen \& Harris(2004)]{van04} VanDalfsen, M.L., \& Harris, W.E.
2004, \aj, 127, 368
\bibitem[VandenBerg et al.(2000)]{van00} VandenBerg, D.A., Swenson, F.J., Rogers, F.J., 
Iglesias, C.A., \& Alexander, D.R. 2000, \apj, 532, 430
\bibitem[Williams et al.(2007)]{will06} Williams, B.F. et al. 2007, \apj, 654, 835  
\bibitem[Worthey et al.(2005)]{wor05} Worthey, G., Espa\~na, A., MacArthur, L., 
\& Courteau, S. 2005, \apj, 631, 820

\end{thebibliography}
\end{document}